\documentclass[12pt]{article}
\pdfoutput=1
\usepackage[top=3cm, bottom=3cm, left=2cm, right=2cm]{geometry}
\usepackage[usenames,dvipsnames,svgnames]{xcolor}
\definecolor{darkblue}{rgb}{0.0,0.1,0.3} 
\definecolor{darkgreen}{rgb}{0,0.65,0}
\definecolor{dblue4}{rgb}{0.06,0.31,0.55} 
\definecolor{nicered}{rgb}{0.7,0.1,0.1}
\definecolor{nicegreen}{rgb}{0.1,0.5,0.1}

\usepackage[numbers,sort&compress]{natbib}

\usepackage[utf8]{inputenc}
\usepackage{amsmath,amssymb,amsfonts,amsthm}
\usepackage{tabularx}
\usepackage{multirow}
\usepackage{dsfont}
\newcolumntype{Y}{>{\centering\arraybackslash}X}

\usepackage{xcolor}
\usepackage[colorlinks=true,citecolor= nicegreen,linkcolor=nicered]{hyperref}

\usepackage{graphicx}
\graphicspath{{figures/}}
\usepackage{cancel}

\usepackage[colorinlistoftodos]{todonotes}

\usepackage{comment}
\includecomment{details}
\specialcomment{details}
{\begingroup}{\endgroup}
\excludecomment{details}

\usepackage{tikz}
\newcommand{\ReportNumbers}[1]{%
\begin{tikzpicture}[overlay, remember picture]
\path (current page.north east) ++(-1,-1) node[below left] {#1};
\end{tikzpicture}
}

\title{Phenomenology of the Zee model for Dirac neutrinos and general neutrino interactions}

\author{Julián Calle\footnote{\href{mailto:julian.callem@udea.edu.co}{julian.callem@udea.edu.co}},
Diego Restrepo\footnote{\href{mailto:restrepo@udea.edu.co}{restrepo@udea.edu.co}}, 
Óscar Zapata\footnote{\href{mailto:oalberto.zapata@udea.edu.co }{oalberto.zapata@udea.edu.co}}\\
\textit{\small Instituto de Física, Universidad de Antioquia, Calle 70 \# 52-21, Apartado Aéreo 1226, Medellín, Colombia}\\
[4mm]
}
\date{\today}
\begin{document}
\maketitle
\ReportNumbers{XXX-XXX}

\begin{abstract}
The Zee model for Dirac neutrinos is one of the simplest models featuring one-loop Dirac neutrino masses.
The interactions between the new scalars (two singly-charged fields) and neutrinos induce general neutrino interactions (GNI) which, as a generalisation of the non standard neutrino interactions, constitute an additional tool to probe models beyond the SM like this.
In this work, we consider a $U(1)_{{B-L}}$ gauge symmetry as the responsible for the Diracness of the neutrinos and the radiative character of the neutrino masses.  
We determine the viable parameter space consistent with neutrino oscillation data, leptonic rare decays and collider constraints, and establish the most relevant experimental prospects regarding  lepton flavor violation searches and GNI in future solar neutrino experiments.
\end{abstract}

\section{Introduction}
\label{sec:intro}
The underlying mechanism behind the massiveness of neutrinos \cite{Fukuda:1998mi,Ahmad:2002jz} holds, after several decades of research dedicated to neutrino physics, as one of the unresolved conundrums of the Standard Model (SM).  In addition to this, the yet-to-be determined question of whether neutrinos are different from their antiparticles \cite{Dolinski:2019nrj} and the lack of signatures of lepton flavor violation (LFV) in the charged sector \cite{Lindner:2016bgg,Calibbi:2017uvl} make the landscape even more unclear. 
Fortunately the very rich experimental program in the lepton sector is undoubtedly rather clear and promising, as for instance, due to the entrance in operation during the current decade of neutrino oscillation experiments such as JUNO~\cite{Djurcic:2015vqa}, DUNE~\cite{Acciarri:2015uup}, Hyper-Kamiokande~\cite{Abe:2018uyc}, among others, and the great experimental effort on searching for LFV with, in some cases, reaching a sensitivity improvement of several orders of magnitude \cite{Baldini:2013ke,Aushev:2010bq,Abrams:2012er,Das:2017nvm,Das:2019fee,Das:2012ze}. 

The simplest way to generate neutrino masses is to add to the SM spectrum three right-handed neutrinos, $\nu_{Ri}$, coupled to SM Higgs through a term in the Lagrangian as~\cite{Minkowski:1977sc,Yanagida:1979as,GellMann:1980vs,Mohapatra:1979ia,Davidson:1978pm}
\begin{align}
  \label{eq:tld}
  \mathcal{L}_{4\nu}=&\,y_D\overline{L}\, \widetilde{H}  \nu_R+\text{h.c.} \,,
\end{align}
with $L$ the lepton doublet, $H$ the SM Higgs doublet and $y_ {D} $ the matrix of neutrino Yukawa couplings. 
Since after the electroweak symmetry breaking (and assuming lepton number conservation) such a term leads to sub-eV neutrino Dirac masses for $|y_D|\lesssim10^{-13}$, the explanation of the smallness of the neutrino mass scale is yet to be settled.  
A way out of this dilemma may be to forbid $\mathcal{L}_{4\nu}$ by imposing a $U(1)_{B-L}$ local symmetry but generate it at loop level via the 5-dimensional operator~\cite{Calle:2018ovc,Bonilla:2018ynb,Saad:2019bqf} 
\begin{align}
  \label{eq:tld}
  \mathcal{L}_{5\nu}=&\,y'_D\overline{L}\, \widetilde{H}  \nu_R \,S+\text{h.c.} \,,
\end{align}
with $S$ being the singlet scalar field responsible for the breaking of the $B-L$ symmetry. From the huge variety of one-loop Dirac neutrino models\footnote{For a recent review see Ref.~\cite{Cai:2017jrq}.} the one including only two singlet charged scalars and three singlet right-handed neutrino~\cite{Nasri:2001ax,Kanemura:2011jj,Jana:2019mez} may be considered as the simplest one -we dub it the Zee model for Dirac neutrinos. 

In this paper we consider the Zee model for Dirac neutrinos with a $B-L$ symmetry, a realization obtained following the approach discussed in Refs.~\cite{Yao:2018ekp, Calle:2018ovc, Bonilla:2018ynb, Jana:2019mgj}. Thanks to this new symmetry (with the charge assignment  given in Table \ref{tab:partcont}) it is possible to forbid the tree level contribution of $\mathcal{L}_{5\nu}$ without any extra {\it ad-hoc} symmetry\footnote{Notice that in Refs. \cite{Nasri:2001ax,Kanemura:2011jj} such contribution is forbidden by a softly-broken $Z_{2}$ symmetry.}. 
As occurs in the original Zee model~\cite{Zee:1980ai, Petcov:1982en}, neutrino masses are generated at one loop with the SM charged leptons running inside the loop, which is closed thanks to a trilinear interaction term between the charged scalars and the Higgs field.  Both charged singlets not only give rise to LFV processes and may leave signatures at the LHC but also induce general neutrino interactions (GNI)~\cite{Bergmann:1999rz, Khan:2019jvr, Bischer:2019ttk}. 
We will determine the viable parameter space consistent with neutrino oscillation observables, LFV and LHC constraints, and we will establish the most relevant experimental perspectives regarding LFV searches and GNI in future solar neutrino experiments. 

The rest of the paper is organized as follows. In section~\ref{sec:Model} we present the model and discuss the neutrino mass generation, the LFV processes and LHC signatures. 
In section~\ref{sec:GNI} we deduce the expressions for the general neutrino interactions, and
the phenomenological analysis is presented in section \ref{sec:results}. Finally, we conclude in section~\ref{sec:conclusions}. 

\section{The Model}
\label{sec:Model}
The realization of the Zee Dirac model through a $U(1)_{{B-L}}$ gauge symmetry and without new extra fermions besides the three right-handed neutrinos $\nu_{Rj}$ requires the introduction of two $SU(2)_L$-singlet charged scalars, $\sigma^{\pm}_{1}$ and $\sigma^{\pm}_{2}$, and the two bosonic fields associated to the new symmetry: $Z'_\mu$ and $S$ as the gauge boson and the scalar responsible for the symmetry breaking, respectively\footnote{If such a breaking is associated to a strong first order phase transition with the subsequent broadcasting of gravitational waves it may be possible to further test this model in the planned gravitational wave experiments (see Refs.~\cite{Jinno:2016knw,Chao:2017ilw,Marzo:2018nov} for related studies).}. The new particle content is summarized in Table~\ref{tab:partcont}, where the charge assignment for the leptons and right-handed neutrinos guarantees a successful anomaly cancellation \cite{Montero:2007cd,Ma:2014qra} and protects the Diracness of neutrinos.

The most general Lagrangian contains the following Yukawa terms 
\begin{align}
\label{Eq:Lag}
-\mathcal{L} &\supset \, f_{\alpha \beta}\overline{L^c_{\alpha}}  L_{\beta} \sigma^{+}_{1} + h_{\beta j} \overline{\ell^c_{R\beta}} \nu_{Rj} \sigma_{2}^{+} + \text{h.c.}\,,
\end{align}
where $L_\alpha$ and $\ell_{R\alpha}$ are lepton doublets and singlets, respectively,  $c$ is charge conjugation operator,  $f_{\alpha\beta}$ and $h_{\beta j}$ are $3 \times 3$  matrices in the flavor space, and $\alpha$, $\beta=1,2,3$ are family indices. 
The coupling $f$ is an antisymmetric complex matrix\footnote{Notice that by performing a field redefinition on $\sigma_1$ it is possible to make one of such couplings real.}
whereas all the non zero $h_{\beta j}$ couplings are complex in general. Note that there are three vanishing couplings  because the lepton number (${L}$) assignment of right-handed neutrino of $L$-charge $-5$: $\nu_{R3}$ $(\nu_{R1})$ for the case of a normal (inverted) hierarchy of neutrino masses\footnote{If only one charged scalar is added  then the right-handed and left-handed neutrinos would have the same charge under the $B-L$ symmetry, thus rendering the Yukawa interaction term $\bar{L}_\alpha\tilde{H}\nu_{Rj}$ allowed.}. This massless chiral fermion, can either contribute to the effective number of relativistic degrees of freedom, $N_{\text{eff}}$, in the early universe~\cite{Saad:2019bqf,Calle:2018ovc}, or becomes massive and be a good Majorana dark matter candidate after the introduction of an extra singlet scalar, $S'$, of $L$-charge $-10$~\cite{Ma:2014qra,Saad:2019bqf}. In the last case, the extra contribution to $N_{\text{eff}}$ comes from the Goldstone boson associated to the imaginary part of the complex singlet scalar field $S'$~\cite{Saad:2019bqf}. 
\begin{table}
  \centering
  \begin{tabular}{|c|c|r|c|}
    \hline  
    Symbol     & $\left( SU(2)_L, U(1)_Y \right)$ & $L$ & \text{Spin}\\ \hline
    $L $  & $(2,-1/2)$ & $-1$ & $1/2$\\
    $H $  & $(2,1/2)$ &  $0$ & $0$\\
    $\overline{ \ell_R } $  & $(1,1)$ & $-1$ & $1/2$\\
    $\overline{ \nu_{R\, i,j}} $  & $(1,0)$ & $4$ & $1/2$\\
    $\overline{ \nu_{R\,k}} $   & $(1,0)$ & $-5$ & $1/2$\\
    $S$ & $(1,0)$ &  $3$ & $0$\\
    $\sigma^{+}_{1}$ & $(1,1)$ & $2$ & $0$\\
    $\sigma^{+}_{2}$ & $(1,1)$ & $5$ & $0$\\ \hline
  \end{tabular}
  \caption{Particle content with the electroweak and $U(1)_{{B-L}}$ quantum numbers. $i=1,j=2,k=3$ ($i=3,j=2,k=1$) correspond to a normal (inverted) neutrino mass ordering. }
  \label{tab:partcont}
\end{table}
The introduction of the extra scalar fields leads to the scalar potential 
\begin{align}\label{eq:V}\nonumber
\mathcal{V}(H,S,\sigma_1,\sigma_2)=&\,\mu_H^{2} H^{\dagger} H + \lambda_H (H^{\dagger} H)^{2} + \mu^{2}_{S} S^{\dagger} S + \lambda_{S} (S^{\dagger} S)^2 + \mu_{1}^{2} |\sigma_{1}^{+}|^{2} + \mu_{2}^{2} |\sigma_{2}^{+}|^{2} \\\nonumber
& + [ \mu_{3}\sigma_{1}^{+} \sigma_{2}^{-} S + \text{h.c.}] + \lambda_{1} |\sigma_{1}^{+}|^{4} + \lambda_{2} |\sigma_{2}^{+}|^{4} + \lambda_{3} (H^{\dagger} H) (S^{\dagger} S) + \lambda_{4} |\sigma_{1}^{+}|^{2} |\sigma_{2}^{+}|^{2}\\
&  + \lambda_{5} (H^{\dagger} H) |\sigma_{1}^{+}|^{2} + \lambda_{6} (H^{\dagger} H) |\sigma_{2}^{+}|^{2}  + \lambda_{7} (S^{\dagger} S) |\sigma_{1}^{+}|^{2} + \lambda_{8} (S^{\dagger} S) |\sigma_{2}^{+}|^{2} \,.
\end{align}
We assume $\mu_H^{2}< 0$ and $\mu_S^{2}< 0$ and $\text{Im}(\mu_3)=0$. Moreover, we assume $\lambda_3\ll1$  such that the scalar $S$ and $H$ do not mix, allowing us to identify the CP even scalar particle in $H$ as the SM Higgs boson. To establish the scalar spectrum, we expand the scalar fields as
\begin{align*}
H = \begin{pmatrix}0 \\ \frac{1}{\sqrt{2}} (h+v_H) \end{pmatrix} \,,&\hspace{1cm}
S = \frac{1}{\sqrt{2}} (S_R+v_{S})\,,
\end{align*}
with $v_H= 246.22$ GeV. Of the original ten scalar degrees of freedom in the model, the gauge bosons $W^{\pm}$, $Z^{0}$ and $Z^{\prime}$ absorb four of them.
Thus, the scalar spectrum contains two neutral CP-even states ($h$ and $S_{R}$) and two charged scalars ($s^{\pm}_{1}$ and $s^{\pm}_{2}$). 
Mass eigenstates for new charged scalars are defined through the mixing angle $\varphi$ as
\begin{align}
\begin{pmatrix}
    s_{1}^{\pm} \\
    s_{2}^{\pm}
\end{pmatrix} = 
\begin{pmatrix}
    \cos \varphi & -\sin \varphi\\
    \sin \varphi & \cos \varphi
\end{pmatrix} 
\begin{pmatrix}
    \sigma_{1}^{\pm} \\
    \sigma_{2}^{\pm}
\end{pmatrix}\,,
\end{align}
where the mixing angle is given by $\sin (2 \varphi) = \sqrt{2}\mu_{3} v_{S} / (m_{s_{2}}^{2}-m_{s_{1}}^{2})$. By defining $m_{\sigma_{1}}^{2} = \mu_{1}^{2}+\lambda_{5}v_{H}^{2}/2 + \lambda_{7}v_{S}^{2}/2$, $m_{\sigma_{2}}^{2} = \mu_{2}^{2}+\lambda_{6}v_{H}^{2}/2+\lambda_{8} v_{S}^{2}/2$, their masses take the form 
\begin{align}
m_{s_{1}, s_{2}}^{2} = \frac{1}{2} \left(m_{\sigma_{2}}^{2} + m_{\sigma_{1}}^{2} \mp \sqrt{(m_{\sigma_{2}}^{2}-m_{\sigma_{1}}^{2})^{2}+8\mu_{3}^{2} v_{S}^{2}}\right)\,,
\end{align}
with $s^\pm_{1}$ been the lightest one. 
Since the effect of $\lambda_{5,6,7,8}$ on $m_{\sigma_i}^2$ can be absorbed by re-scaling the $\mu_i^2$ parameters, hereafter we further assume $\lambda_{5,6,7,8}\ll1$.

\subsection{Radiative neutrino masses}\label{sec:Neutrinos}
\begin{figure}
\centering
\includegraphics[scale=0.6]{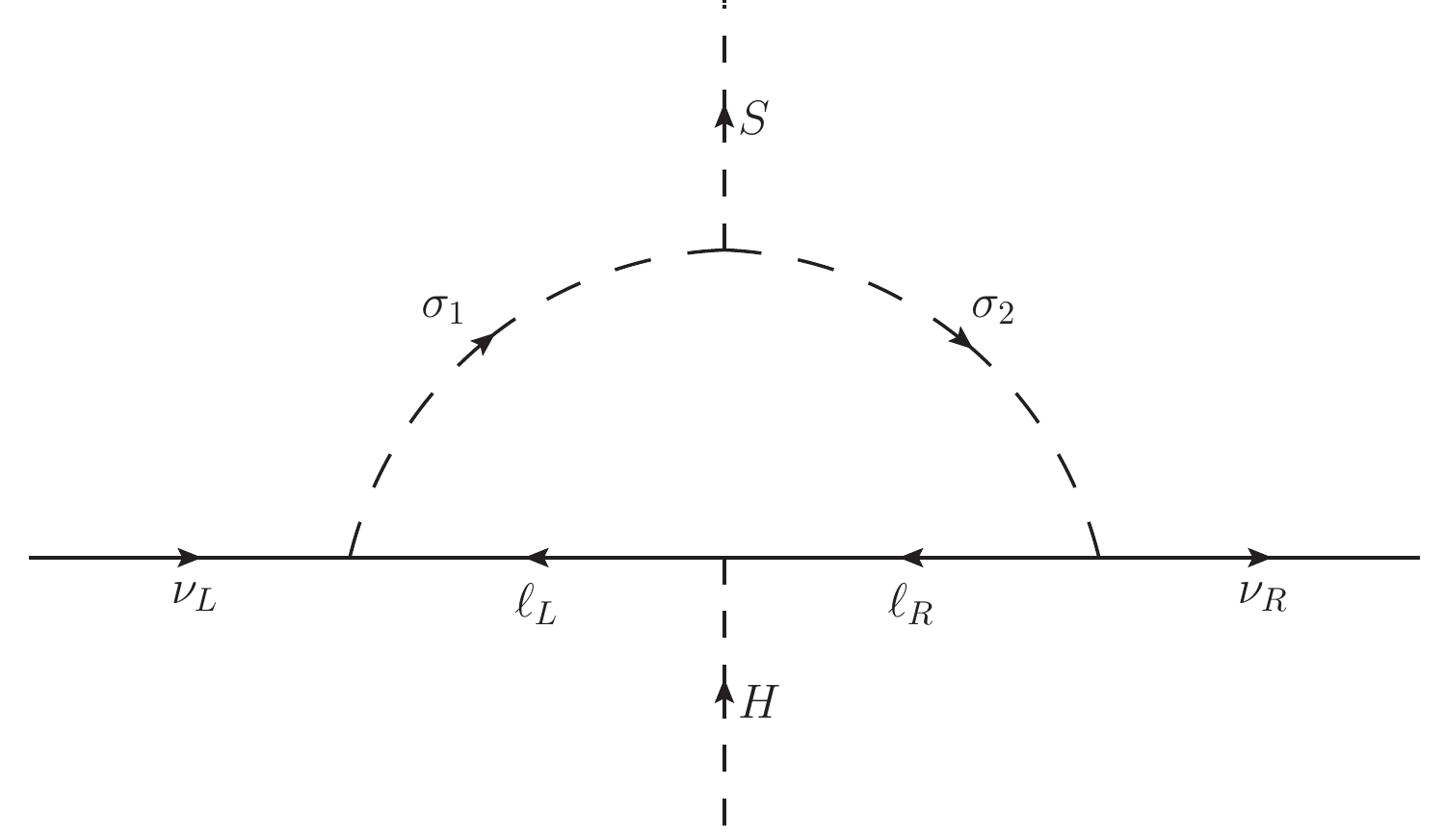}
\caption{B-L flux in the one-loop diagram leading to Dirac neutrino masses.}
\label{fig:zee}
\end{figure}
The interplay of the new Yukawa interactions and the trilinear interaction $\mu_3$ generates Dirac neutrino masses at one loop level as displayed in Fig.~\ref{fig:zee}. The expression for the corresponding mass matrix can be cast as
\begin{align}
\label{eq:Mnu}
\left[M_{\nu}\right]_{\alpha i} = \kappa [f^{T}]_{\alpha \beta} [M_{\ell}]_{\beta\beta} h_{\beta i},
\end{align}
where $M_{\ell}$ is the diagonal mass matrix of the charged leptons and
\begin{align}
\kappa = \frac{\sin (2\varphi)}{16 \pi^{2}} \ln \frac{m_{s_{2}}^{2}}{m_{s_{1}}^{2}}\,.
\end{align}
$M_{\nu}$ is diagonalized by the biunitary transformation 
\begin{align}
\left(U^{(\nu)}_L\right)^{\dagger} M_{\nu} U^{(\nu)}_R = M_{\nu}^{\text{diag}} = \text{diag}(m_{\nu_1}, m_{\nu_2}, m_{\nu_3})\,,
\end{align}
where $U^{(\nu)}_R$ and $U^{(\nu)}_L$  are rotation $3 \times 3$ matrices associated to the right-handed and left-handed neutrinos,  respectively.
Assuming $U_{R}^{(\nu)} =\mathds{1}$ the lepton mixing matrix takes the form  
\begin{align}
U_{\text{PMNS}} = U^{(\nu)}_L\,.
\end{align}
Due to the antisymmetric character of the Yukawa coupling matrix $f$ one neutrino state remains massless, and from Eq.~(\ref{eq:Mnu}) one can express six of the nine non-zero Yukawa-couplings in terms of the neutrino oscillation observables, the masses of the charged leptons and scalars. The expressions for the non-zero couplings in the case of a normal neutrino mass hierarchy are:
\begin{align}
    h_{22}&= -\frac{\left(U_{13} U_{32}-U_{12} U_{33}\right) \left(f_{13} h_{12} \kappa  m_e-m_2 U_{32}\right)}{f_{13}
   \kappa  \left(U_{23} U_{32}-U_{22} U_{33}\right) m_{\mu }},\\
   h_{23}&= \frac{\left(U_{12} U_{33}-U_{13} U_{32}\right)
   \left(f_{13} h_{13} \kappa  m_e-m_3 U_{33}\right)}{f_{13} \kappa  \left(U_{23} U_{32}-U_{22} U_{33}\right) m_{\mu
   }},\\
   h_{32}&= \frac{f_{13} h_{12} \kappa m_e( U_{13} U_{22} - U_{12} U_{23})+m_2 U_{22} (-U_{13} U_{32}
   +U_{12} U_{33}) }{f_{13} \kappa  \left(U_{23} U_{32}-U_{22} U_{33}\right) m_{\tau }},\\
   h_{33}&= \frac{f_{13}
   h_{13} \kappa m_e (U_{13} U_{22} - U_{12} U_{23}) + m_3 U_{23} (-U_{13}U_{32}+U_{12}
   U_{33})}{f_{13} \kappa  \left(U_{23} U_{32}-U_{22} U_{33}\right) m_{\tau }},\\
   f_{12}&= \frac{f_{13} \left(U_{13} U_{22}-U_{12}
   U_{23}\right)}{U_{13} U_{32}-U_{12} U_{33}},\\
   f_{23}&= \frac{f_{13} \left(U_{23} U_{32}-U_{22} U_{33}\right)}{U_{13}
   U_{32}-U_{12} U_{33}},
\end{align}
with $U_{ij}=(U_{\text{PMNS}})_{ij}$, $m_2=\sqrt{\Delta m^2_{\mathrm{sol}}}$ and $m_3=\sqrt{\Delta m^2_{\mathrm{atm}}}$. 
The expressions for the case of an inverted neutrino mass hierarchy (IH) are:
\begin{align}
    h_{22}&= \frac{\left(U_{11} U_{32}-U_{12} U_{31}\right) \left(f_{13} h_{12} \kappa  m_e-m_2 U_{32}\right)}{f_{13}
   \kappa  \left(U_{22} U_{31}-U_{21} U_{32}\right) m_{\mu }},\\
   h_{32}&= \frac{f_{13} h_{12} \kappa m_e (U_{12} U_{21}-U_{11} U_{22}) + m_2 U_{22} ( -U_{12} U_{31}+ U_{11} U_{32})}{f_{13} \kappa  \left(U_{22}
   U_{31}-U_{21} U_{32}\right) m_{\tau }},\\
   h_{21}&= -\frac{\left(U_{12} U_{31}-U_{11} U_{32}\right) \left(f_{13} h_{11} \kappa 
   m_e-m_1 U_{31}\right)}{f_{13} \kappa  \left(U_{22} U_{31}-U_{21} U_{32}\right) m_{\mu }},\\
   h_{31}&= \frac{f_{13} h_{11}
   \kappa m_e (U_{12} U_{21} - U_{11} U_{22} ) + m_1 U_{21} (-U_{12} U_{31} + U_{11} U_{32})}{f_{13}
   \kappa  \left(U_{22} U_{31}-U_{21} U_{32}\right) m_{\tau }},\\
   f_{12}&= \frac{f_{13} \left(U_{12} U_{21}-U_{11}
   U_{22}\right)}{U_{12} U_{31}-U_{11} U_{32}},\\
   f_{23}&= \frac{f_{13} \left(U_{22} U_{31}-U_{21} U_{32}\right)}{U_{12}
   U_{31}-U_{11} U_{32}},
\end{align}
with $m_1=\sqrt{\Delta m^2_{\mathrm{atm}}}$ and $m_2=\sqrt{\Delta m^2_{\mathrm{sol }}+\Delta m^2_{\mathrm{atm}}}$.

Inserting the best fit values \cite{deSalas:2020pgw} for the mixing angles and squared mass differences along with $\delta_{\text{CP}}=\pi (0)$ for a NH (IH), the non-free $f_{\alpha\beta}$ couplings show a mild hierarchy,
\begin{align}
  &    f_{12}\approx1.81f_{13},\,\,\,f_{23}\approx2.93f_{13},\hspace{1cm}\text{for NH};\\
  &    f_{12}\approx-0.88f_{13},\,\,\,f_{23}\approx-0.20f_{13},\hspace{1cm}\text{for IH}.
\end{align}
On the other hand, from the expressions for the non-free $h_{\beta i}$ Yukawa couplings it is highly expected that those associated to the tau ($h_{3i}$) turn to be suppressed in comparison to the ones associated to the muon ($h_{2i}$), and these in turn suppressed in comparison to the free $h_{\beta i}$ parameters ($h_{12}$ and $h_{13}$ for NH and $h_{11}$ and $h_{12}$ for IH).  
Introducing the benchmark point $m_{s1}=m_{s2}/2=300$ GeV, $\sin(2\varphi)=0.5$, $f_{13}=0.01$ and $h_{12}=h_{13}=0.1\,(h_{11}=h_{12}=0.1)$ along the previous inputs, the non-free $h_{\beta i}$ couplings become
\begin{align}
&h_{22}\approx -0.00027,\,\,h_{23}\approx -0.00027,\,\,h_{32}\approx 0.00001,\,\,h_{33}\approx 0.00001,\nonumber\\
&f_{12}\approx 0.00738,\,\,f_{23}\approx 0.01761,\hspace{1cm}\text{for NH};\\
&h_{22}\approx 0.00243,\,\,h_{32}\approx 0.00013,\,\,h_{21}\approx 0.00238,\,\,h_{31}\approx 0.00012,\nonumber\\
&f_{12}\approx -0.00876,\,\,f_{23}\approx-0.00202,\hspace{1cm}\text{for IH}.
\end{align}
These results illustrate the strong hierarchy present in the $h_{\beta i}$ couplings. 

Here a comment is in order concerning the leptonic CP phase, $\delta_{\text{CP}}$. Since only one $f_{ij}$ Yukawa coupling -say $f_{13}$- can be rendered real, it follows from the expression for $f_{12}$ and $f_{23}$ that the CP symmetry is conserved in the neutrino sector only in the scenario when those couplings are real. That is, such a scenario demands that the Dirac phase necessarily takes the values $\delta_{\text{CP}}=0,\pi$. This scenario is in agreement to a large extent with the most recent analysis \cite{Esteban:2020cvm,deSalas:2020pgw} which give a best fit for the complex phase close to $\delta_{\text{CP}}=\pi$ for a normal hierarchy (NH), whereas  $\delta_{\text{CP}}=0$ is marginally allowed within the 3$\sigma$ range for a inverted hierarchy (IH).   
This in turn makes such a scenario to be falsifiable at the upcoming neutrino experiments such as DUNE \cite{Acciarri:2015uup} and Hyper-Kamiokande \cite{Abe:2018uyc}.

\subsection{LFV processes} \label{sec:LFV}
%
\begin{figure}
\centering
\includegraphics[scale=0.6]{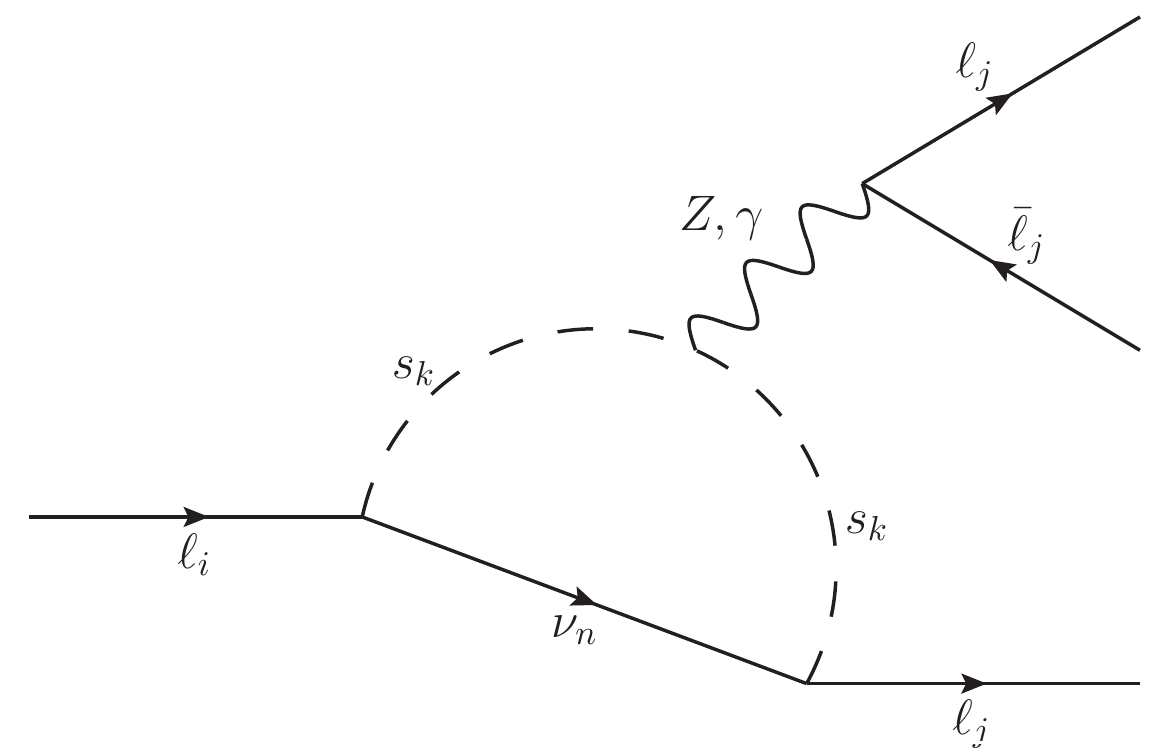} \hspace{1cm}
\includegraphics[scale=0.6]{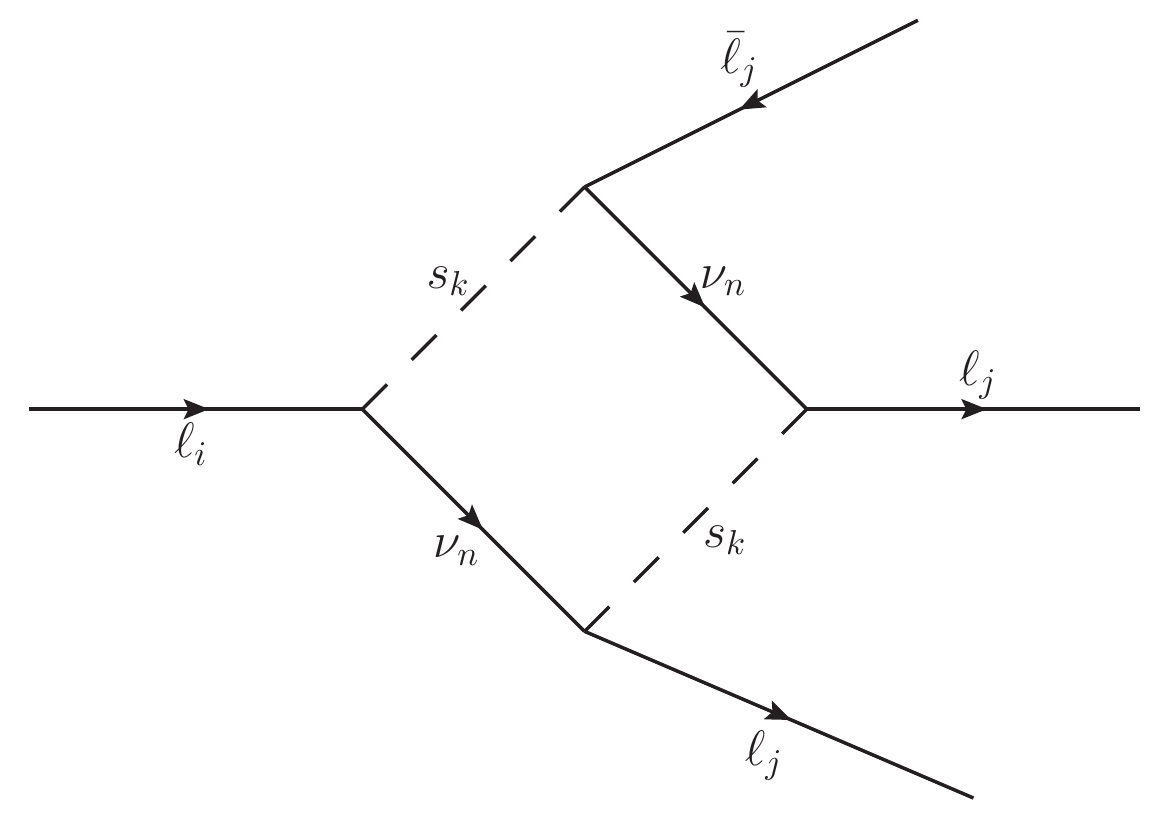}
\caption{One-loop Feynman diagrams for the rare decays $\ell_{i} \to \ell_{j} \bar{\ell}_{j} \ell_{j}$.}
\label{fig:LFVdiagrams}
\end{figure}
The same Yukawa interactions involved in the neutrino mass generation also induce charged lepton flavor violating processes such as $\ell_{i} \rightarrow \ell_{j}\gamma$, $\ell_{i} \rightarrow 3\ell_{j}$ and $\mu-e$ conversion in nuclei. 
In this model such processes are generated at one-loop level and are mediated by the charged scalars $s_k^\pm$ and neutrinos. The expressions for the branching ratios $\mathcal{B}(\ell_i\to\ell_j\gamma)$ are given by  
\begin{align}
\mathcal{B}(\mu\to e\gamma)&=\frac{\alpha_{e}Br(\mu\to e\nu_\mu\bar{\nu}_e)}{768\pi  G_F^2}\left[ 16 C_{\varphi12}^2|f_{13}|^2|f_{23}|^2 +  C_{\varphi21}^2(|h_{12} h_{22}+h_{13} h_{23}|^2) \right],\\
\mathcal{B}(\tau\to e\gamma)&=\frac{\alpha_{e}Br(\tau\to e\nu_\tau\bar{\nu}_e)}{768\pi  G_F^2}\left[ 16 C_{\varphi12}^2|f_{12}|^2|f_{23}|^2 +  C_{\varphi21}^2(|h_{12}h_{32}+h_{13}h_{33}|^2 ) \right],\\
\mathcal{B}(\tau\to \mu\gamma)&=\frac{\alpha_{e}Br(\tau\to \mu\nu_\tau\bar{\nu}_\mu)}{768\pi  G_F^2}\left[ 16 C_{\varphi12}^2|f_{12}|^2|f_{13}|^2 +  C_{\varphi21}^2(|h_{22}h_{32}+h_{23}h_{33}|^2) \right],
\end{align}
for the case of a NH, and 
\begin{align}
\mathcal{B}(\mu\to e\gamma)&=\frac{\alpha_{e}Br(\mu\to e\nu_\mu\bar{\nu}_e)}{768\pi  G_F^2}\left[ 16 C_{\varphi12}^2|f_{13}|^2|f_{23}|^2 +  C_{\varphi21}^2(|h_{11}h_{21}+h_{12}h_{22}|^2) \right],\\
\mathcal{B}(\tau\to e\gamma)&=\frac{\alpha_{e}Br(\tau\to e\nu_\tau\bar{\nu}_e)}{768\pi  G_F^2}\left[ 16 C_{\varphi12}^2|f_{12}|^2|f_{23}|^2 +  C_{\varphi21}^2(|h_{11}h_{31}+h_{12}h_{32}|^2 ) \right],\\
\mathcal{B}(\tau\to \mu\gamma)&=\frac{\alpha_{e}Br(\tau\to \mu\nu_\tau\bar{\nu}_\mu)}{768\pi  G_F^2}\left[ 16 C_{\varphi12}^2|f_{12}|^2|f_{13}|^2 +  C_{\varphi21}^2(|h_{21}h_{31}+h_{22}h_{32}|^2) \right],
\end{align}
for a IH. Here we have defined $C_{\varphi ij}=\cos\varphi^2/m^2_{si}+\sin\varphi^2/m^2_{sj}$.   
Notice that, in contrast to neutrino masses, these branching ratios are not suppressed by the scalar mixing angle $\varphi$, as is expected since the rare decays $\ell_{i} \rightarrow \ell_{j} \gamma$ do not depend on the massiveness of neutrinos. 

As regards to the $\ell_i \rightarrow \ell_j\bar{\ell_j}\ell_j$ processes there exist two main diagrams (see Fig.~\ref{fig:LFVdiagrams}) contributing to the total amplitude: the photon- and $Z$- penguin diagrams (left panel) and the box diagram (right panel).   
The contribution of the Higgs-penguin diagram is subdominant due to the suppression coming from the Yukawa couplings associated to the first two families of charged leptons (the contribution of the processes involving tau leptons is not negligible but the corresponding limits are less restrictive) whereas the $Z'$-penguin contribution is also subdominant due to the large mass of the $Z'$ gauge boson required to surpass the LHC lower bound (see next section). 
The $\mu-e$ conversion diagrams are obtained when the pair of lepton lines attached to the photon and $Z$ boson in the penguin diagrams are replaced by a pair of light quark lines\footnote{Higgs-penguin diagrams are again suppressed, in this case by the Yukawa couplings of light quarks.}. 
The corresponding amplitudes for the $\ell_i \rightarrow \ell_j\bar{\ell}_j\ell_j$, $\ell_i \rightarrow \ell_j\bar{\ell}_k\ell_k$ and $\mu-e$ processes are calculated by using the {\tt FlavorKit} code \cite{Porod:2014xia}.

Finally, when performing our numerical analysis in Section \ref{sec:results} we will consider the current experimental bounds and their future expectations shown in Table~\ref{tab:LFVbounds}. 
\begin{table}[t!]
\begin{center}
\begin{tabular}{|cc|cc|c|}
\hline 
Observable & & Present limit & & Future sensitivity \\
\hline\hline 
$\mathcal{B}(\mu \to e\gamma)$ & & $4.2 \times 10^{-13}$ & & $ 6.3 \times 10^{-14}$  \\
$\mathcal{B}(\tau \to e\gamma)$&&$3.3 \times 10^{-8}$ && $3\times 10^{-9}$ \\
$\mathcal{B}(\tau\to\mu\gamma)$&& $4.4 \times 10^{-8}$ && $3\times 10^{-9}$ \\
$\mathcal{B}(\mu \to eee)$ & & $ 1.0\times 10^{-12}$  & &$   10^{-16}$   \\  
$\mathcal{B}(\tau \to eee)$ & & $ 2.7\times 10^{-8}$ & &$  3\times 10^{-9}$   \\  
$\mathcal{B}(\tau \to \mu\mu\mu)$ & & $ 2.1\times 10^{-8}$  & &$   10^{-9}$   \\  
$\mathcal{B}(\tau \to e \mu\mu)$ & & $ 2.7 \times 10^{-8}$ & & $-$  \\  
$\mathcal{B}(\tau \to \mu e e)$ & & $ 1.8\times 10^{-8}$  & & $-$ \\  
${\rm R_{\mu e}({\rm Ti})}$ & & $4.3 \times 10^{-12}$ & & $  10^{-18}$ \\
${\rm R_{\mu e}({\rm Au})}$ & & $7.3 \times 10^{-13}$ & &  $-$ \\
\hline
\end{tabular}
\end{center}
\caption{ Current bounds \cite{TheMEG:2016wtm,Aubert:2009ag,Bona:2007qt,Miyazaki:2013yaa,Bellgardt:1987du,Hayasaka:2010np,Dohmen:1993mp,Bertl:2006up} and projected sensitivities \cite{Baldini:2013ke,Aushev:2010bq,Blondel:2013ia,Abrams:2012er} for the charged LFV observables.}
\label{tab:LFVbounds}
\end{table}

\subsection{LHC observables}\label{sec:coll}

LHC searches for additional neutral gauge bosons have delivered stringent constraints on the fraction of the $Z'$ mass to the gauge coupling, $M_{Z'}/g'$.  The recasting~\cite{Chiang:2019ajm} of the latest ATLAS results for the search of dilepton resonances using $139\ \text{fb}^{-1}$~\cite{Aad:2019fac} for the $U(1)_{{B-L}}$ model gives $  M_{Z'}\gtrsim 5$ TeV for $g'\sim 0.1$. Let us recall that the mass of the $Z'$ boson is given by $M_{Z'}=3g'v_S$, which in turn implies $v_S\gtrsim 17$ TeV for $g'\sim 0.1$.     

Regarding collider bounds on the charged scalar $s_1^\pm$, its main production mechanism  at the LHC for small and intermediate Yukawa couplings ($f_{\alpha\beta},h_{\beta i}\lesssim 0.1$) is via Drell-Yan processes. The subsequent decays into a lepton and a neutrino involve precisely these Yukawa interactions\footnote{Notice that the decays involving gauge bosons are not present.}. 
  
The expressions for decay width of the processes $\Gamma(s_1\to \ell\sum_i\nu_i)\equiv \Gamma(s_1\to \ell\nu)$ for NH are given by 
\begin{align}
\Gamma(s_1\to e\nu)&=\,\frac{m_{s1}}{16\pi}\left(4\cos\varphi^2(|f_{12}|^2+|f_{13}|^2)+\sin\varphi^2(|h_{12}|^2+|h_{13}|^2)\right),\\
\Gamma(s_1\to \mu\nu)&=\,\frac{m_{s1}}{16\pi}\left(4\cos\varphi^2(|f_{12}|^2+|f_{23}|^2)+\sin\varphi^2(|h_{22}|^2+|h_{23}|^2)\right)\\
\Gamma(s_1\to \tau\nu)&=\,\frac{m_{s1}}{16\pi}\left(4\cos\varphi^2(|f_{13}|^2+|f_{23}|^2)+\sin\varphi^2(|h_{32}|^2+|h_{33}|^2)\right),
\end{align}
whereas for IH they read
\begin{align}
\Gamma(s_1\to e\nu)&=\,\frac{m_{s1}}{16\pi}\left(4\cos\varphi^2(|f_{12}|^2+|f_{13}|^2)+\sin\varphi^2(|h_{11}|^2+|h_{12}|^2)\right),\\
\Gamma(s_1\to \mu\nu)&=\,\frac{m_{s1}}{16\pi}\left(4\cos\varphi^2(|f_{12}|^2+|f_{23}|^2)+\sin\varphi^2(|h_{21}|^2+|h_{22}|^2)\right),\\
\Gamma(s_1\to \tau\nu)&=\,\frac{m_{s1}}{16\pi}\left(4\cos\varphi^2(|f_{13}|^2+|f_{23}|^2)+\sin\varphi^2(|h_{31}|^2+|h_{32}|^2)\right).
\end{align}
These direct leptonic decays lead to the collider signature of dileptons plus missing transverse momentum since
\begin{align}
    pp\rightarrow \gamma^*/Z^*\rightarrow s_1^+s_1^-\rightarrow \ell_i^+\ell_j^-\nu\nu.
\end{align}
Such a signature is analogous to the one of electroweak production of sleptons in the context of simplified supersymmetric scenarios \cite{Aad:2019vnb,Sirunyan:2020eab}. 
Assuming a 100\% branching ratio into $\ell=e, \mu$ and an integrated luminosity of $139\,\text{fb}^{-1}$ ATLAS has set upper limits on the slepton pair production cross sections in such a way right-handed slepton masses are excluded up to 420 GeV for vanishing neutralino masses \cite{Aad:2019vnb}. 
A recast of the excluded cross section for slepton pair production was done in Ref.~\cite{Crivellin:2020klg}, which allows to exclude $SU(2)_L$-singlet charged scalars decaying into a lepton plus a neutrino with masses up to 200 GeV (or even more in some cases). Moreover, the analysis in Ref.~\cite{Alcaide:2019kdr}\footnote{See Refs.~\cite{Cao:2017ffm,Babu:2019mfe} for related works.} shows that singly charged scalars with masses below 500 GeV decaying mostly into electrons and muons may be excluded at the high luminosity phase of the LHC. 

On the other hand, single production of charged scalars can also take place (for large Yukawa couplings) through the radiation from a lepton external leg in $s$-channel diagrams featuring a gauge boson ($\gamma, Z$ or $W$). The collider signatures for these topologies are one, two, or three leptons plus missing transverse momentum. Concerning to the final state with three leptons ($e$ or $\mu$), a recasting \cite{Alcaide:2017dcx} of the ATLAS trilepton searches~\cite{Aad:2014hja} excludes charged scalar not decaying into taus with masses $\sim190$ GeV and Yukawa couplings of $\mathcal{O}(1)$.  

\section{General neutrino interactions in Zee Dirac model}
\label{sec:GNI}
\begin{table}[t!]
\begin{center}
\begin{tabular}{|c|c|c|}
\hline 
GNI & Current bound & Projected sensitivity \\
\hline\hline       
$\epsilon_{ee}^V$ & $[-0.12,0.08]$ & $[-0.016,0.016]$ \\
$\epsilon_{ee}^A$ & $[-0.13,0.07]$ & $[-0.016,0.016]$ \\
$\tilde{\epsilon}_{ee}^V$ & $[-0.07,0.13]$ & $[-0.016,0.016]$ \\
$\tilde{\epsilon}_{ee}^A$ & $[-0.08,0.13]$ & $[-0.016,0.016]$ \\
$\epsilon_{\mu\mu}^V/\epsilon_{\tau\tau}^V$ & $[-0.22,0.08]$ & $[-0.1,0.1]$ \\
$\epsilon_{\mu\mu}^A/\epsilon_{\tau\tau}^A$ & $[-0.08,0.08]$ & $[-0.03,0.03]$ \\
$\tilde{\epsilon}_{\mu\mu}^V/\tilde{\epsilon}_{\tau\tau}^V$ & $[-0.09,0.08]$ & $[-0.04,0.04]$ \\
$\tilde{\epsilon}_{\mu\mu}^A/\tilde{\epsilon}_{\tau\tau}^A$ & $[-0.90,0.22]$ & $[-0.1,0.1]$ \\
$\epsilon_{\mu\mu}^S/\epsilon_{\tau\tau}^S$ & $[-0.83,0.83]$ & $[-0.5,0.5]$ \\
$\epsilon_{\mu\mu}^P/\epsilon_{\tau\tau}^P$ & $[-0.83,0.83]$ & $[-1.22,1.22]$ \\
$\epsilon_{\mu\mu}^T/\epsilon_{\tau\tau}^T$ & $[-0.15,0.15]$ & $[-0.1,0.1]$ \\
\hline
\end{tabular}
\end{center}
\caption{ Current bounds and projected sensitivities for GNI \cite{Khan:2019jvr}.}
\label{tab:GNIbounds}
\end{table}

The $h_{\beta i}$ and $f_{\alpha\beta}$ interactions also induce new effective four-fermion interactions between neutrinos and charged leptons with all the Lorentz-invariant structures: scalar ($S$), pseudoscalar ($P$), vector ($V$), axialvector ($A$) and tensor ($T$)\footnote{There are also effective four-fermion interactions mediated by the $Z'$ boson, however, they turn to be heavily suppressed by $M_{Z'}$}. 
These new interactions can modify the SM prospect for the neutrino-electron elastic scattering, and therefore be constrained \cite{Rodejohann:2017vup,Bischer:2019ttk,Khan:2019jvr} by using the solar neutrinos experiment results such as Borexino~\cite{Bellini:2011rx}, TEXONO~\cite{Deniz:2009mu} and CHARM-II~\cite{Vilain:1994qy}.      

The most general effective Lagrangian that describes the GNI can be cast as\footnote{We closely follow the notation introduced in Ref.~\cite{Khan:2019jvr}}
\begin{align}
    \label{eq:Lag_GNI}
    \mathcal{L} = \frac{G_{F}}{\sqrt{2}} \sum_{a = S, P, V, A, T} \left( \overline{\nu}_{\alpha} \Gamma^{a} \nu_{\beta} \right) \left[ \overline{\ell} \Gamma^{a} \left( \epsilon^{a}_{\alpha \beta} + \Tilde{\epsilon}^{a}_{\alpha \beta} i^{a} \gamma^{5} \right) \ell \right]\,,
\end{align}
where $\{\Gamma^S,\Gamma^P,\Gamma^V,\Gamma^A,\Gamma^T\} \equiv \{I,i\gamma^5,\gamma^\mu,\gamma^\mu\gamma^5,\sigma^{\mu\nu}\}$, $\{i^S,i^P,i^V,i^A,i^T\}\equiv\{i,i,1,1,i\}$ and $G_{F}$ is the Fermi constant. 
The  $3\times3$ matrices $\epsilon$ and $\tilde{\epsilon}$ -which are assumed to be real-  parametrize the departures of the SM result. 

On the other hand, the effective four-fermion Lagrangian for GNI contributions in the present model is given by
\begin{align}
    \mathcal{L}_{\text{eff}} =& -2f_{\alpha \rho} f^{*}_{\beta \sigma} [(\overline{\ell_\beta}\gamma^\mu P_L\ell_\alpha)(\overline{\nu_\sigma}\gamma_\mu P_L\nu_\rho)]C_{\phi 12}-\frac{1}{2}h_{\alpha \rho} h^{*}_{\beta \sigma} [(\overline{\ell_\beta}\gamma^\mu P_R\ell_\alpha)(\overline{\nu_\sigma}\gamma_\mu P_R\nu_\rho)]C_{\phi 21}\nonumber\\
    &+f_{\alpha \rho} h^{*}_{\beta \sigma} \left[(\overline{\ell_\beta}P_L\ell_\alpha)(\overline{\nu_\sigma}P_L\nu_\rho) - \frac{1}{4}( \overline{\ell_\beta} \sigma^{\mu\nu} P_L \ell_\alpha ) ( \overline{\nu_\sigma} \sigma_{\mu\nu} P_L \nu_\rho )\right]D_{\phi12}\nonumber\\     &+f_{\beta \sigma}^{*} h_{\alpha \rho} \left[(\overline{\ell_\beta}P_R\ell_\alpha)(\overline{\nu_\sigma}P_R\nu_\rho) - \frac{1}{4}( \overline{\ell_\beta} \sigma^{\mu\nu} P_R \ell_\alpha ) ( \overline{\nu_\sigma} \sigma_{\mu\nu} P_R \nu_\rho )\right]D_{\phi12},
\end{align}
where $D_{\varphi ij}=\cos\varphi\sin\varphi(1/m^2_{sj}-1/m^2_{si})$.   
It follows that the factors in the second and third lines in the previous expression become suppressed for a small scalar mixing angle, while the two factors in the first line are suppressed for small Yukawa couplings correspondingly.  

\begin{figure}[t]
\centering
\includegraphics[scale=0.4]{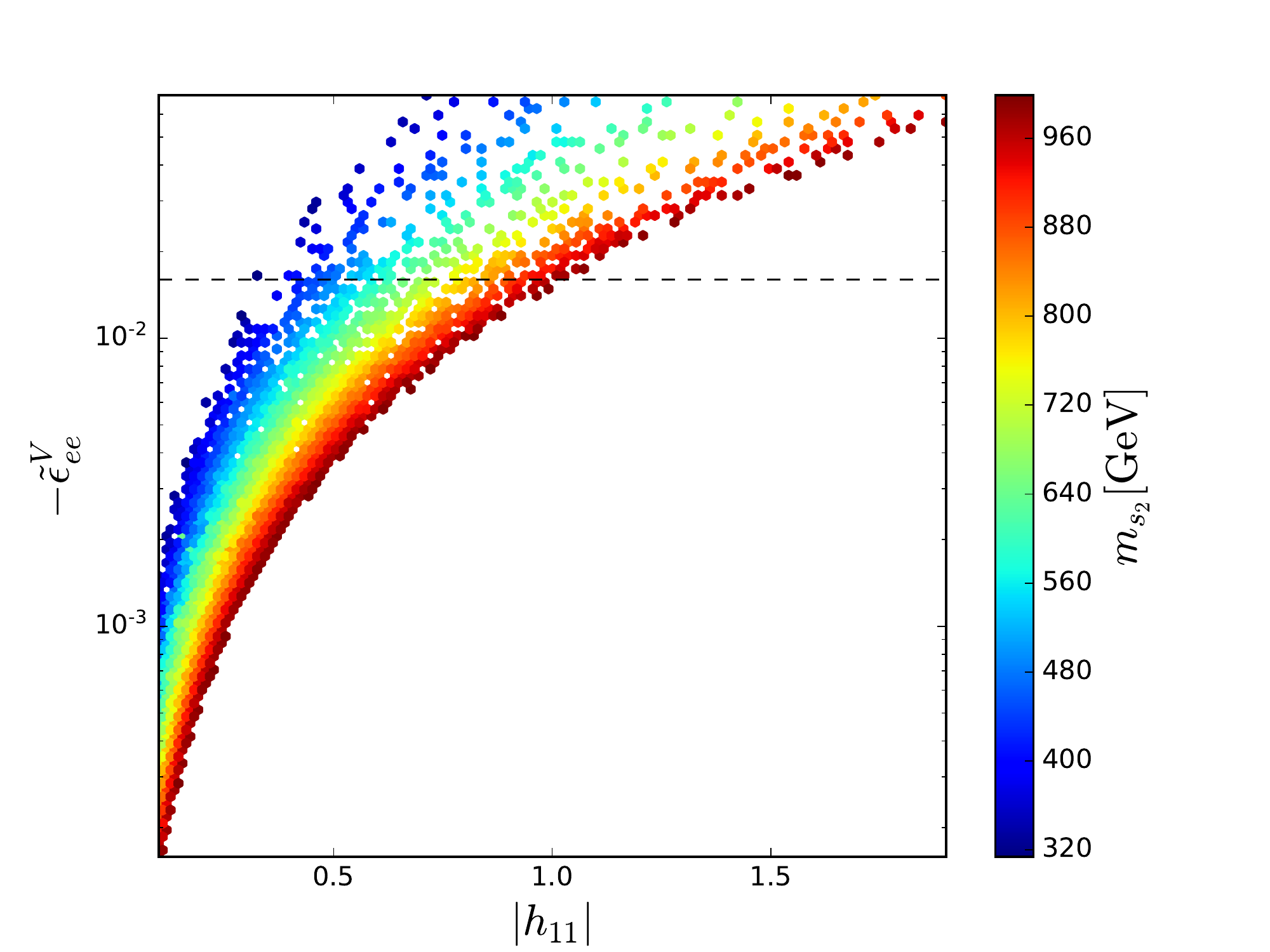}
\includegraphics[scale=0.4]{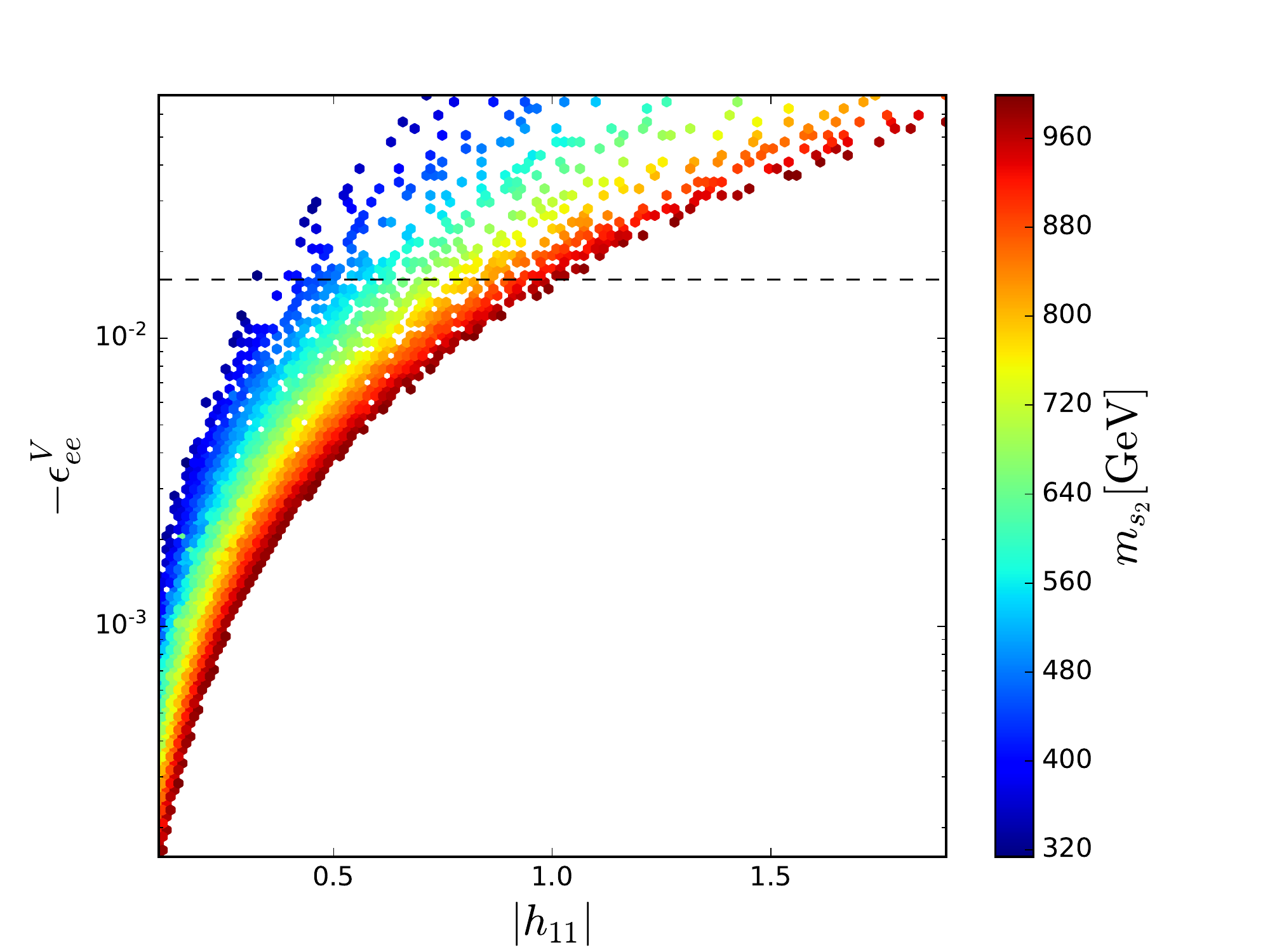}\\
\includegraphics[scale=0.4]{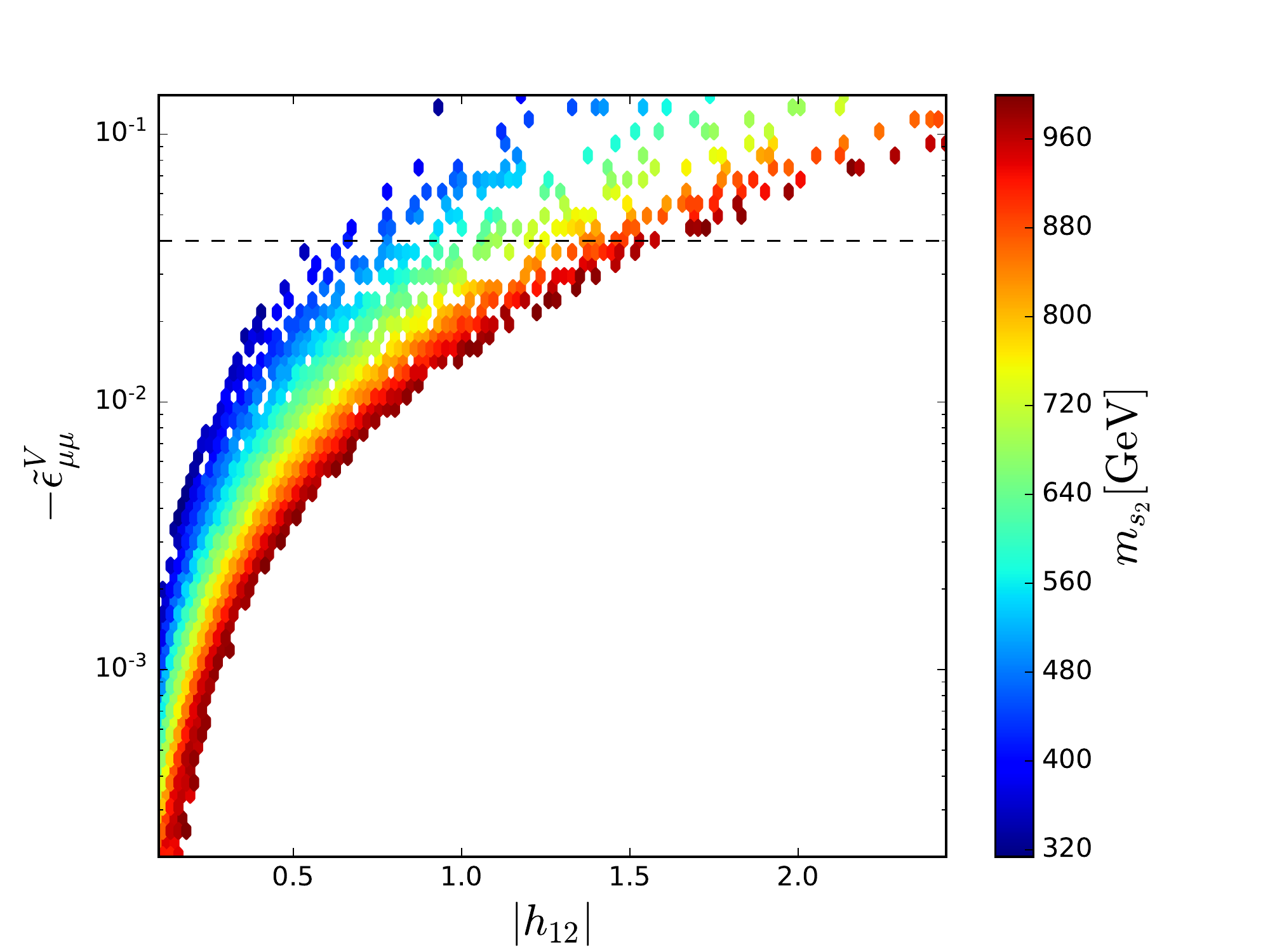}
\includegraphics[scale=0.4]{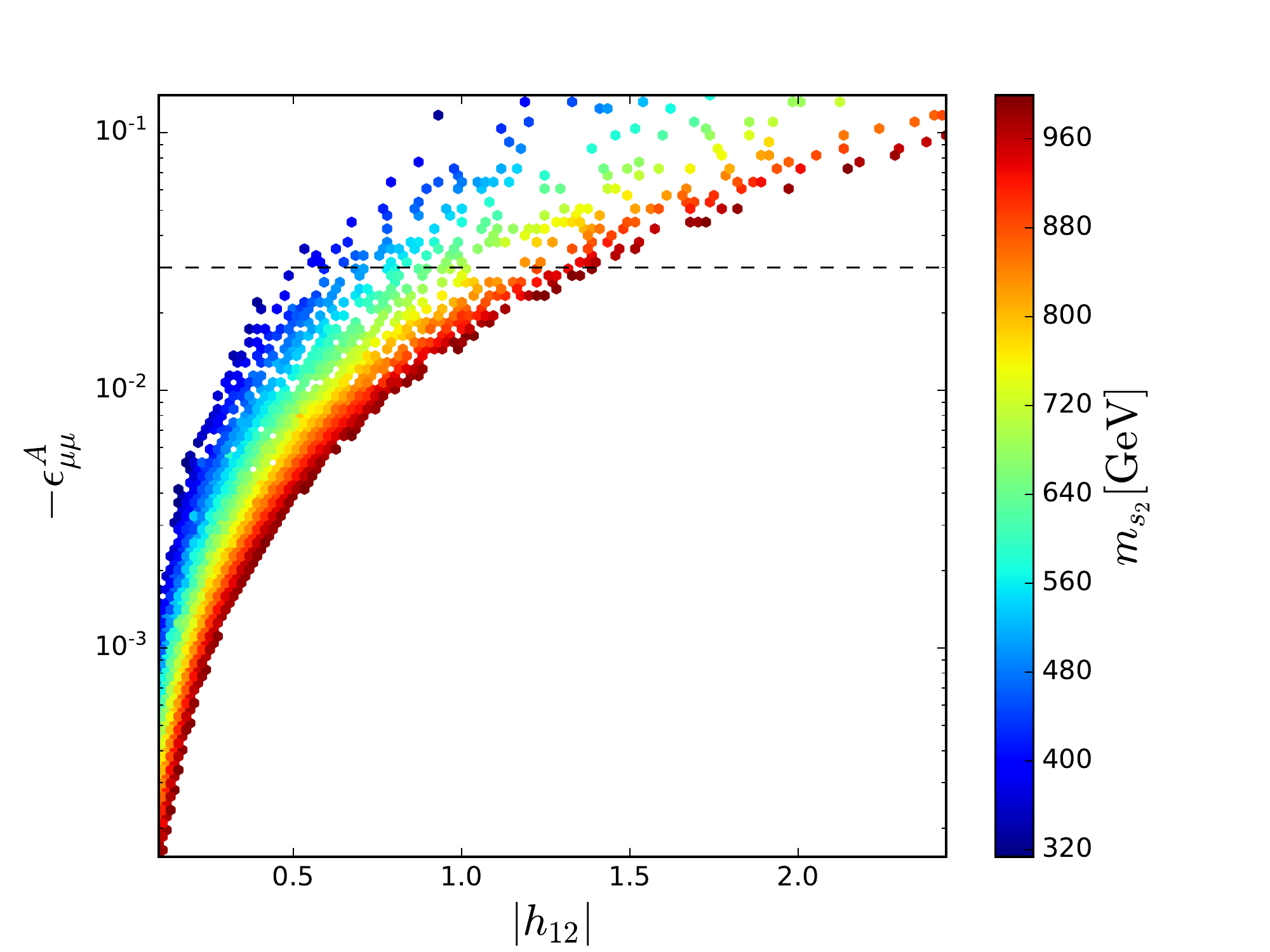}
\caption{General neutrino interactions for the IH case. }
\label{fig:GNI-IH}
\end{figure}

For electron neutrinos $\sigma=\rho=e$ the corresponding GNI read
\begin{align}
  &\epsilon^{V}_{ee}=\tilde{\epsilon}^V_{ee}=\epsilon^{A}_{ee}=\tilde{\epsilon}^A_{ee}=-\frac{\sqrt{2}}{8G_F}|h_{11}|^2C_{\varphi 21},\\
  &\epsilon^{S}_{ee}=\tilde{\epsilon}^S_{ee}=\epsilon^{P}_{ee}=\tilde{\epsilon}^P_{ee}=\epsilon^{T}_{ee}=\tilde{\epsilon}^T_{ee}=0.
\end{align}
Notice that all GNI associated to electron neutrinos for the NH are zero since $h_{i1}=0$.
For muon neutrinos $\sigma=\rho=\mu$ it follows that 
\begin{align}
  &\epsilon^{V}_{\mu\mu}=\epsilon^{A}_{\mu\mu}=-\frac{\sqrt{2}}{8G_F}(|h_{12}|^2C_{\phi 21} +4|f_{12}|^2C_{\varphi 12}),\\
  &\tilde{\epsilon}^{V}_{\mu\mu}=\tilde{\epsilon}^{A}_{\mu\mu}=-\frac{\sqrt{2}}{8G_F}(|h_{12}|^2C_{\varphi 21} -4|f_{12}|^2C_{\varphi 12}),\\
  &\epsilon^{S}_{\mu\mu}=-\epsilon^{P}_{\mu\mu}=-4\epsilon^{T}_{\mu\mu}=\frac{\sqrt{2}D_{\varphi12}}{4G_F}(f_{12} h^{*}_{12} +f_{12}^{*} h_{12}),\\
  &\tilde{\epsilon}^S_{\mu\mu}=-\tilde{\epsilon}^P_{\mu\mu}=-4\tilde{\epsilon}^T_{\mu\mu}=\frac{i\sqrt{2}D_{\phi12}}{4G_F}(f_{12} h^{*}_{12} - f_{12}^{*} h_{12}).
\end{align}
For the case of tau neutrinos $\sigma=\rho=\tau$ the GNI are 
\begin{align}
  &\epsilon^{V}_{\tau\tau}=\epsilon^{A}_{\tau\tau}=-\frac{\sqrt{2}}{8G_F}(|h_{13}|^2C_{\varphi 21} +4|f_{13}|^2C_{\varphi 12}),\\
  &\tilde{\epsilon}^{V}_{\tau\tau}=\tilde{\epsilon}^{A}_{\tau\tau}=-\frac{\sqrt{2}}{8G_F}(|h_{13}|^2C_{\varphi 21} -4|f_{13}|^2C_{\varphi 12}),\\
  &\epsilon^{S}_{\tau\tau}=-\epsilon^{P}_{\tau\tau}=-4\epsilon^{T}_{\tau\tau}=\frac{\sqrt{2}D_{\varphi12}}{4G_F}(f_{13} h^{*}_{13} +f_{13}^{*} h_{13}),\\
  &\tilde{\epsilon}^S_{\tau\tau}=-\tilde{\epsilon}^P_{\tau\tau}=-4\tilde{\epsilon}^T_{\tau\tau}=\frac{i\sqrt{2}D_{\phi12}}{4G_F}(f_{13} h^{*}_{13} - f_{13}^{*} h_{13}).
\end{align}
Because $h_{i3}=0$ in a inverted neutrino mass ordering  these GNI reduce to
\begin{align}
  &\epsilon^{V}_{\tau\tau}=\epsilon^{A}_{\tau\tau}=-\tilde{\epsilon}^{V}_{\tau\tau}=-\tilde{\epsilon}^{A}_{\tau\tau}=-\frac{\sqrt{2}}{2G_F}|f_{13}|^2C_{\varphi 12},\\
  &\epsilon^{S}_{\tau\tau}=\epsilon^{P}_{\tau\tau}=\epsilon^{T}_{\tau\tau}=\tilde{\epsilon}^S_{\tau\tau}=\tilde{\epsilon}^P_{\tau\tau}=\tilde{\epsilon}^T_{\tau\tau}=0.
\end{align}

By using the Borexino results, the authors of Ref.~\cite{Khan:2019jvr} have derived  90\% C.L. constraints on general neutrino interactions for all neutrino flavors. The most constraining bounds that apply to the current model are displayed in Table \ref{tab:GNIbounds}.  

\section{Results and discussion}\label{sec:results}
\begin{figure}[t]
\centering
\includegraphics[scale=0.4]{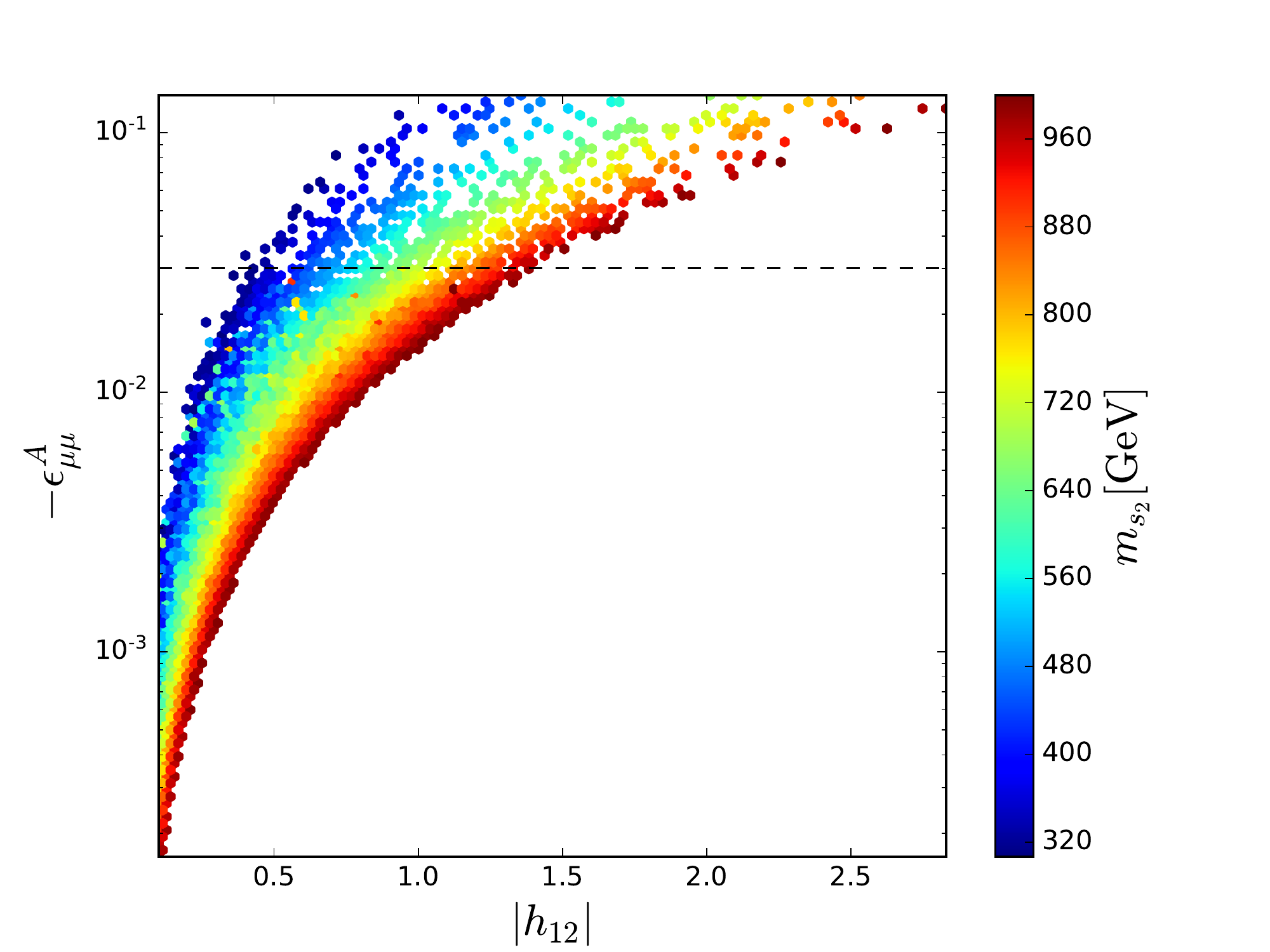}
\includegraphics[scale=0.4]{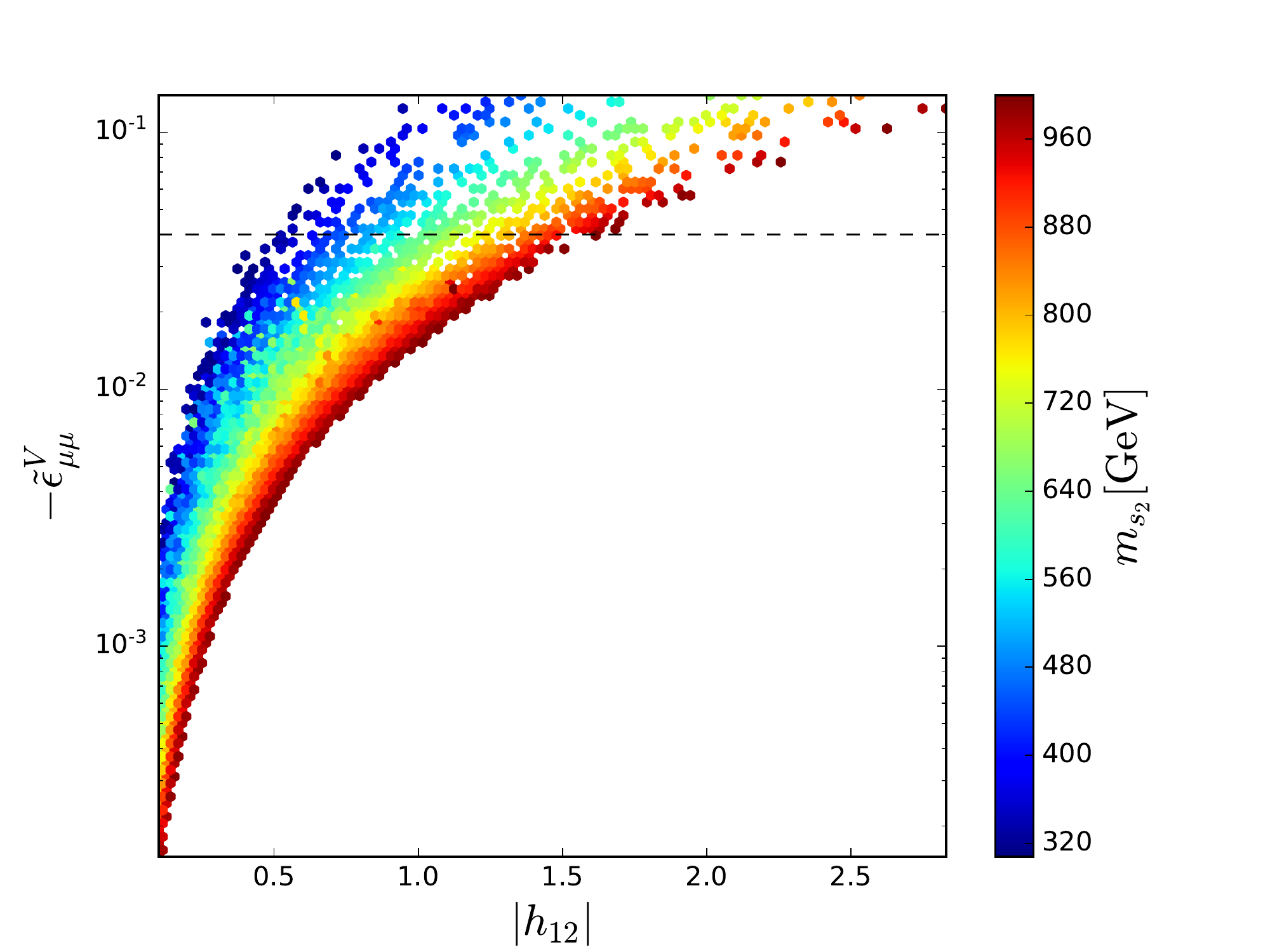}\\
\includegraphics[scale=0.4]{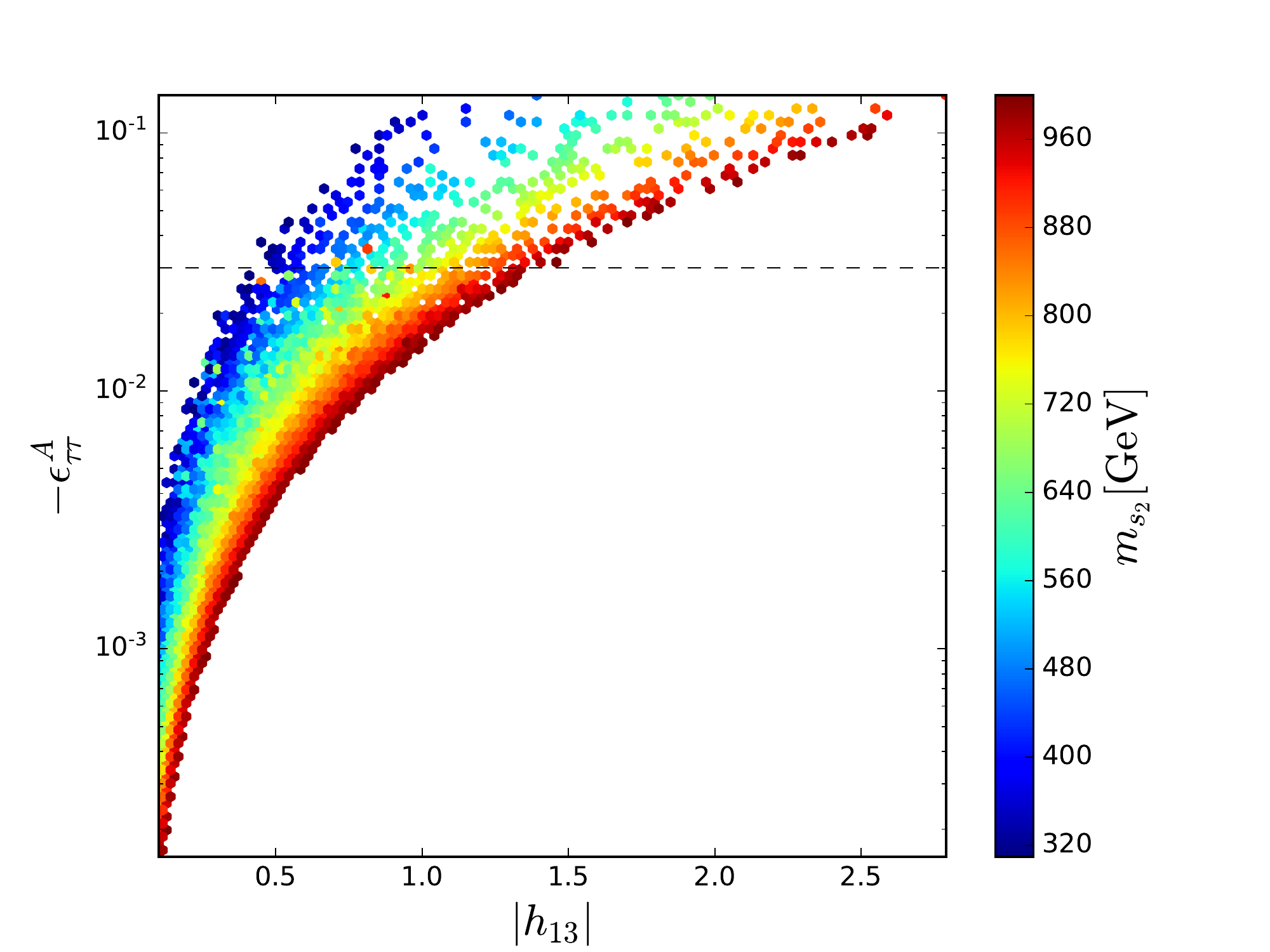}
\includegraphics[scale=0.4]{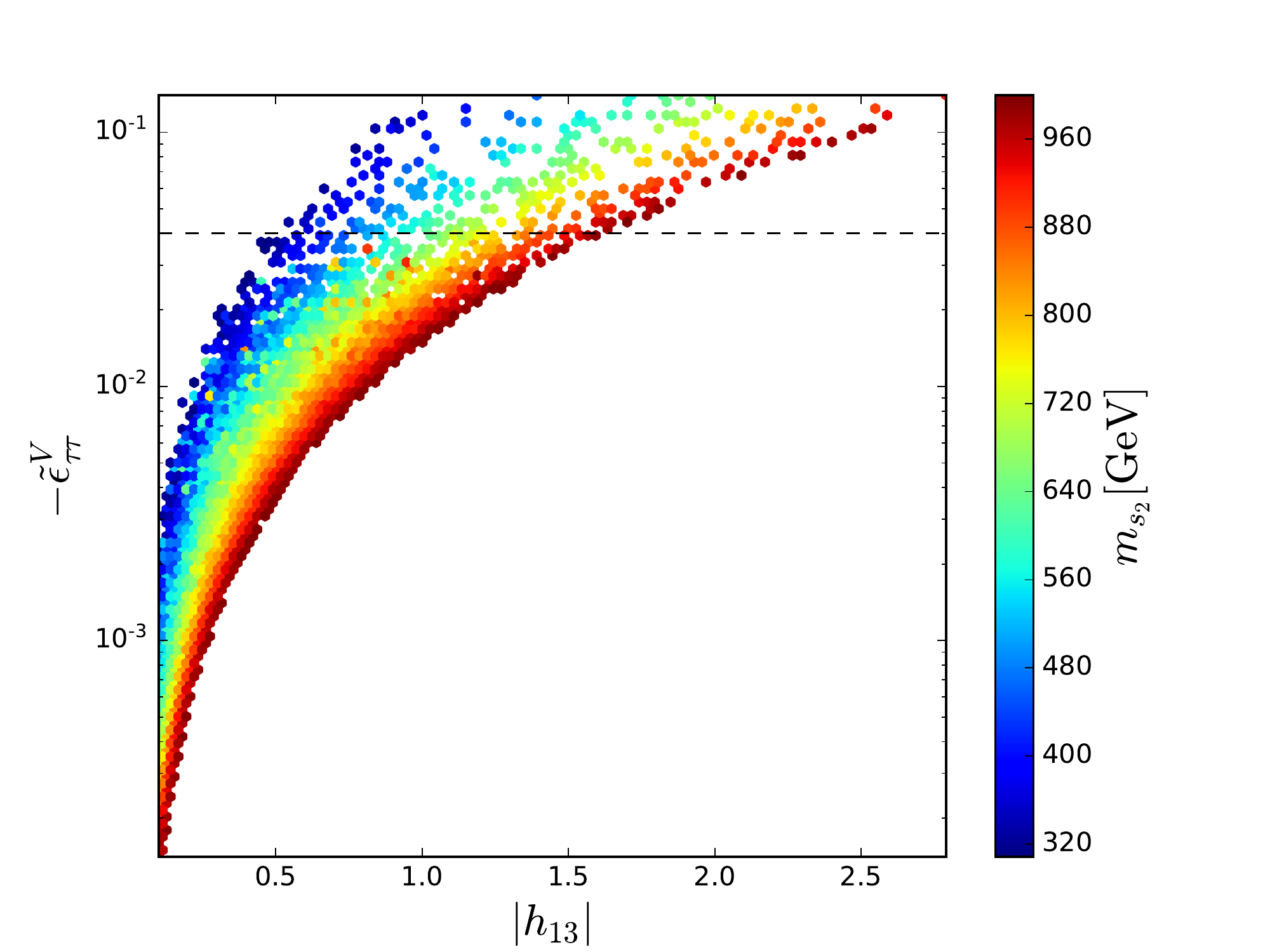}
\caption{General neutrino interactions for the NH case.}
\label{fig:GNI-NH}
\end{figure}
We have performed a random scan for each neutrino mass ordering, varying just a subset of the free parameters of the model with the aim of making the analysis simpler. 
The set of free parameters of the model relevant for our analysis  has been varied as
\begin{align}
  \label{eq:scanLFVlow2}
  & 10^{-5} \leq |f_{13}|\leq 3\;;\;\nonumber\\
  & 0.1 \leq |h_{12}, h_{13} (h_{11}, h_{12})|\leq 3\;;\;\nonumber\\
  &10^{-6} \leq \varphi \leq \pi/2\;;\;    \nonumber\\
  & 80 \, {\rm GeV} \leq m_{s1}\leq 500\, {\rm GeV}\; ; m_{s_2} =  [m_{s_1},1000\,{\rm GeV}].
\end{align}
The magnitude of the non-free Yukawa couplings are restricted to be within the range $[10^{-5},3]$, and the other parameters of the scalar potential were fixed as $\lambda_i=10^{-4},\,i=1,...,8$. The parameters of the new gauge boson are set to $M_ {Z'} = 6$~TeV and $g' = 0.5$ in such a way $\upsilon_{S} = 4$~TeV.\footnote{With these values it follows that $\mu_3<200$ GeV, in order to obtain charged scalars below the TeV mass scale.} 
In our numerical analysis, all the viable benchmark points satisfy the current neutrino oscillation data within the $3 \sigma$ level~\cite{Esteban:2020cvm}, both for inverted and normal hierarchies, with $\delta_{\text{CP}}=\pi (0)$ for a NH (IH). 
Likewise, they satisfy the LFV upper bounds reported in Table~\ref{tab:LFVbounds} and the GNI constraints in Table~\ref{tab:GNIbounds}. 
Finally, we impose the collider  bounds  on  the  charged  scalar coming from searches for final states with an oppositely charged lepton pair ($ e^{-} e^{+} $ and $ \mu^{-} \mu^{+} $ ) and missing transverse energy~\cite{Crivellin:2020klg}.
In order to obtain the particle spectrum and low energy observables we have used {\tt SPheno} \cite{Porod:2003um,Porod:2011nf} and the {\tt FlavorKit} \cite{Porod:2014xia} of {\tt SARAH} \cite{Staub:2013tta,Staub:2015kfa}.

\begin{figure}[t]
\centering
\includegraphics[scale=0.4]{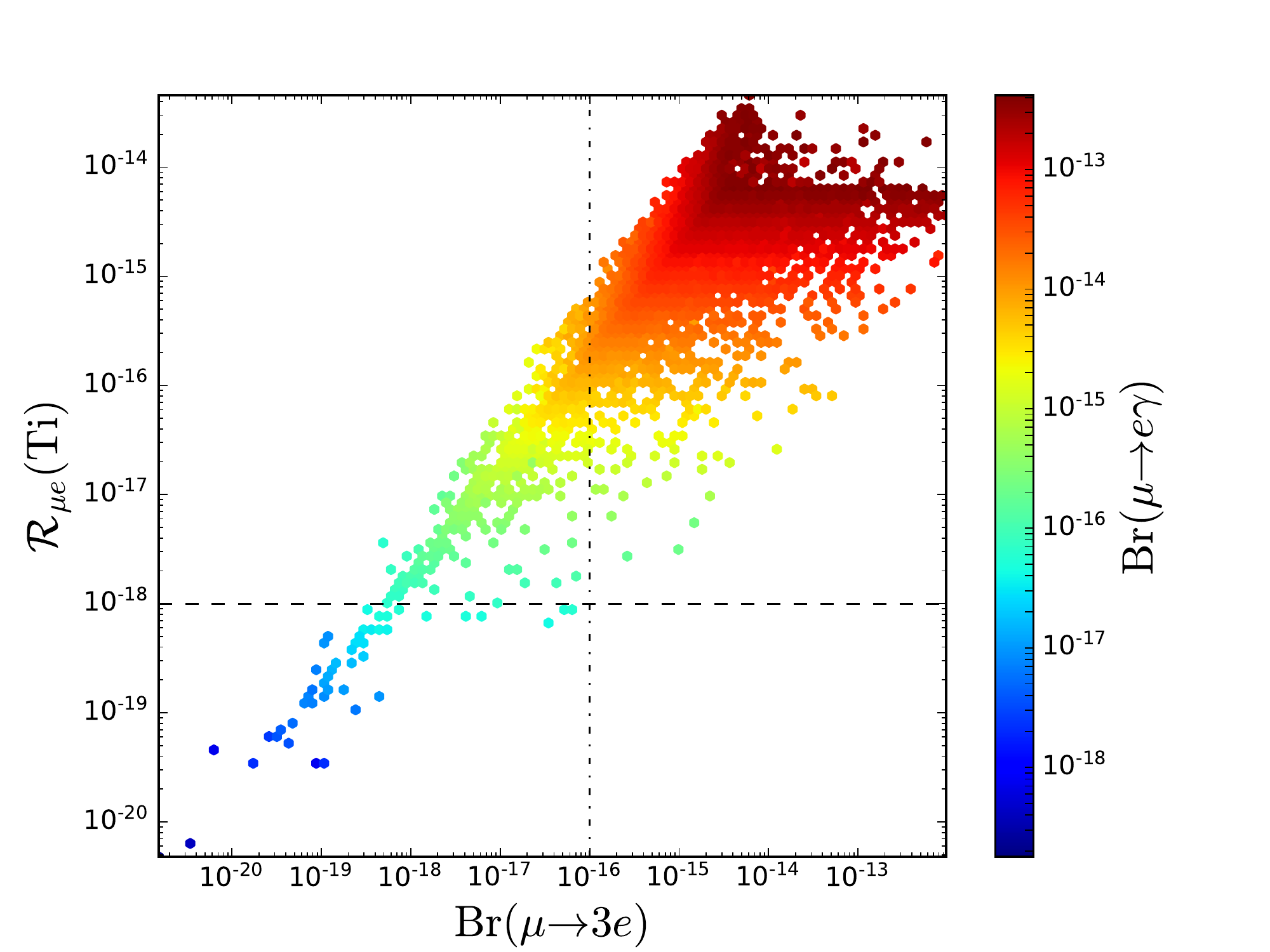}
\includegraphics[scale=0.4]{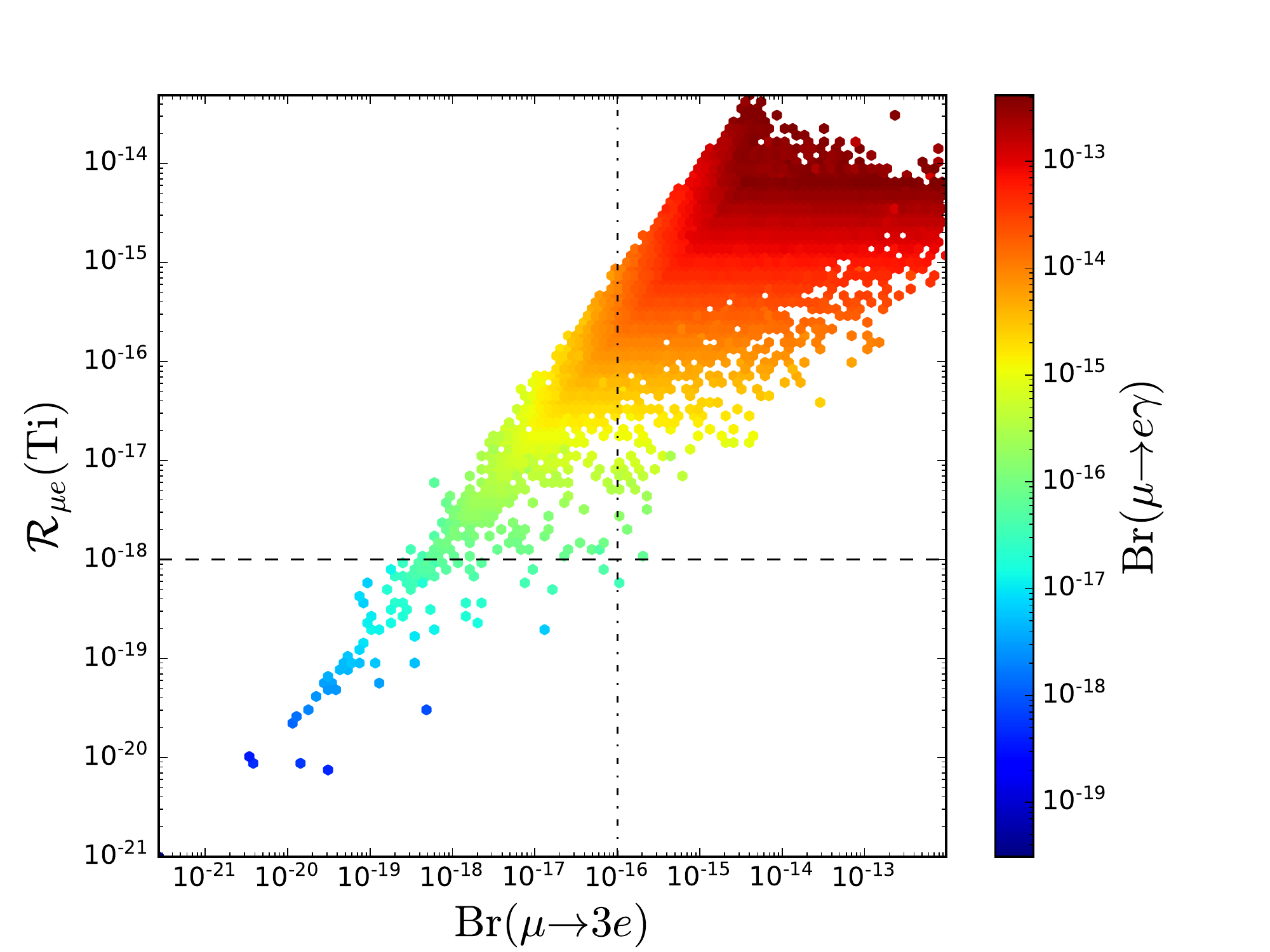}\\
\caption{LFV processes for IH (left) and NH (right). The dotted and dashed lines represent the sensitivity limit expected for the future searches for $\mathcal{R}_{\mu e} ({\rm Ti})$ and $\mathcal{B}(\mu \to 3e)$
, respectively.}
\label{fig:LFV}
\end{figure}
\begin{figure}[t]
\centering
\includegraphics[scale=0.4]{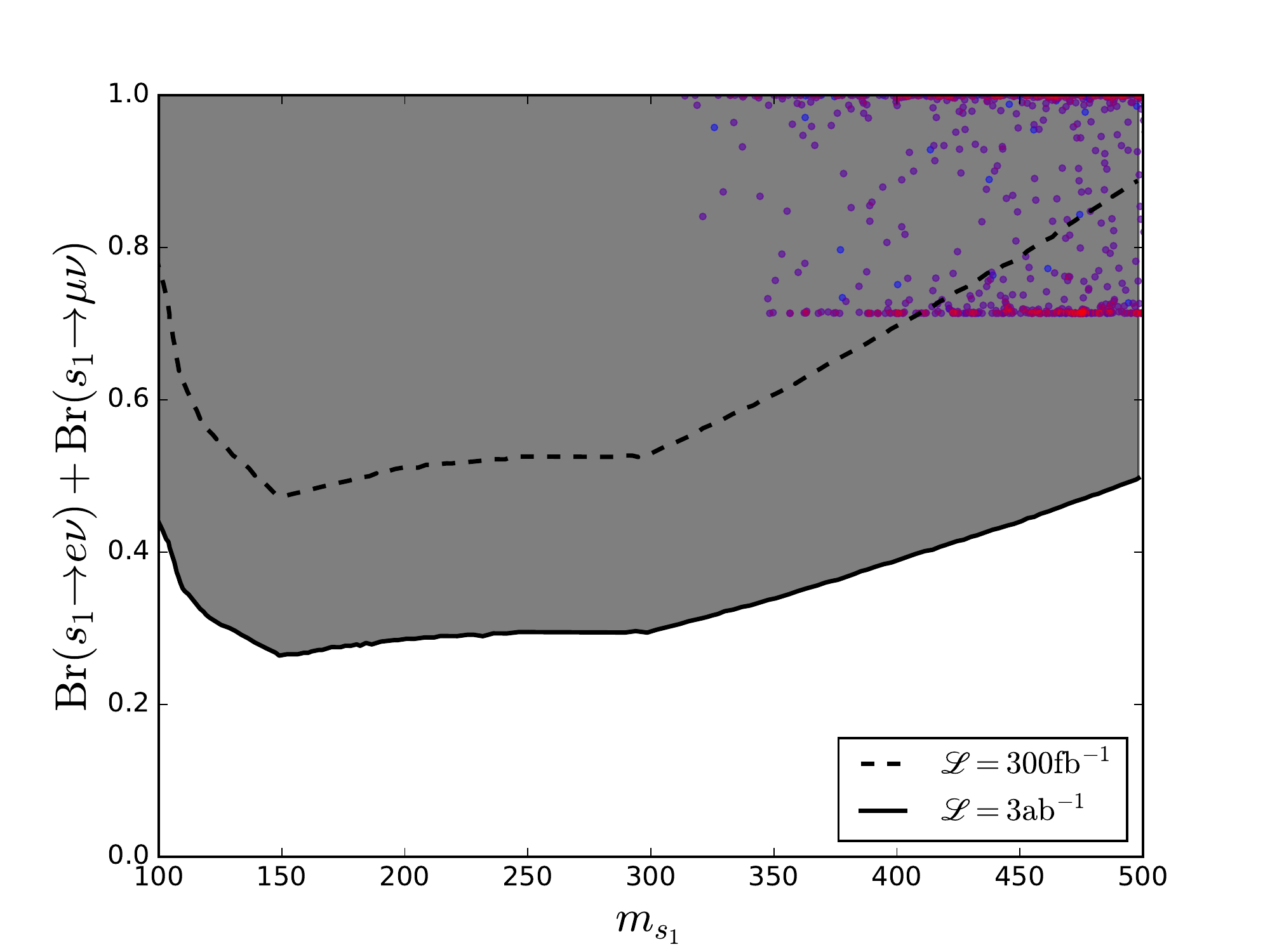}
\includegraphics[scale=0.4]{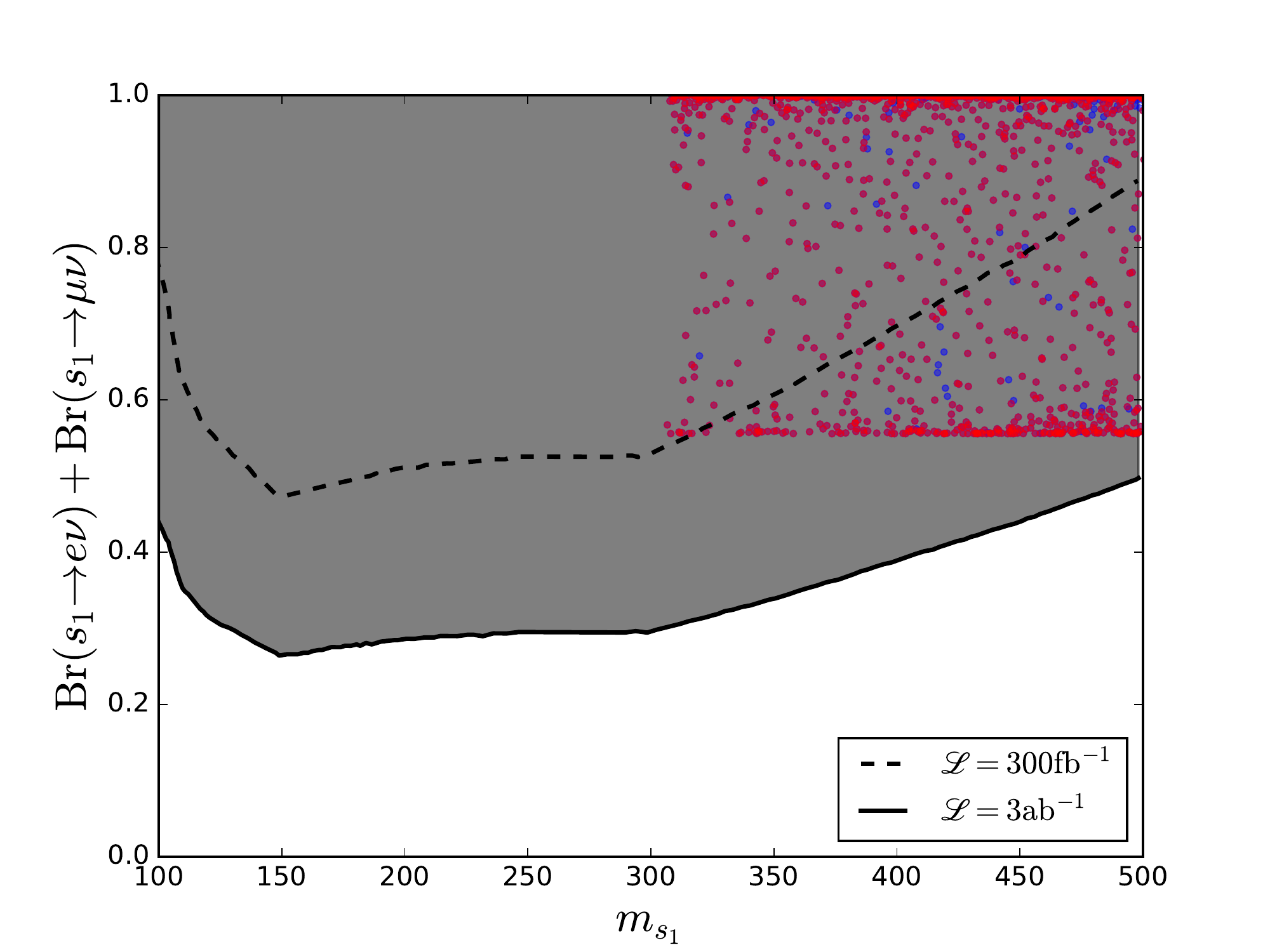}
\caption{Projected exclusion reach in the decay branching ratio of $s_1^{\pm}$ into electrons and muons with luminosities of 300 fb$^{-1}$ (dashed line) and 3 ab$^{-1}$ (solid line) for IH (left) and NH (right). 
The red dots are within the future sensitivity of the LFV experiments (see Fig.~\ref{fig:LFV}).}
\label{fig:Charged}
\end{figure}

In the Fig.~\ref{fig:GNI-IH} (\ref{fig:GNI-NH}) are displayed the results for the general neutrino interactions  assuming a inverted (normal) neutrino mass hierarchy, and the future sensitivity (dashed line) that may be reached in future solar neutrino experiments~\cite{Khan:2019jvr}, with the color bar denoting the mass of the heaviest charged scalar. 
Only the results for the parameters $\tilde{\epsilon}^{V}_{ee}$, $\tilde{\epsilon}^{V}_{\mu \mu}$, $\epsilon^{V}_{ee}$ and $\epsilon^{A}_{\mu \mu}$ ($\tilde{\epsilon}^{V}_{\mu \mu}$, $\tilde{\epsilon}^{V}_{\tau \tau}$, $\epsilon^{A}_{\mu \mu}$ and $\epsilon^{A}_{\tau \tau}$) are displayed since they are the ones with the most appreciable restrictions on future experiments. 
As is expected, the heavier the $s_2^\pm$, the lower the intensity of the general neutrino interactions.     

The correlation between the most constrained LFV processes, namely $\mathcal{B}(\mu \to e \gamma)$, $\mathcal{B}(\mu \to e \gamma)$ and $\mu-e$ conversion, is shown in the top panels of the Fig.~\ref{fig:LFV}. The corresponding prospects for the future sensitivities (dashed and dotted lines) point out that a large portion of the parameter space will be explored.
On the other hand, the current limits impose mild restrictions on the Yukawa couplings, in particular $\mathcal{B}(\mu \to e \gamma) $ establishes that $f_{13}$ is restricted to values less than $0.04$, furthermore, future searches for these processes restrict the value of $f_{13}$ around to $10^{-3}$.
Finally, the Fig.~\ref{fig:Charged} shows the results of the search for charged scalars in the LHC. The current limits impose that the mass of the lightest scalar must be greater than $\sim 260$ GeV and $\sim 300$ GeV for normal and inverse hierarchy, respectively \cite{Crivellin:2020klg}. The dashed line shows the future exclusion limits for an integrated luminosity of 300fb$^{-1}$ and the solid line for 3ab$^{-1}$~\cite{Alcaide:2019kdr}. 
Notice that in principle it is possible to explore the whole parameter space in the high luminosity phase of the LHC.

Since the relevant GNI parameters and the decay rates for $s_1\to\ell\nu$ depend only on the magnitude of the Yukawa couplings it is understandable to expect that the above results hold in the scenario where the Yukawa couplings are complex. 
In order to check this we have performed an additional scan for the scenario where the Dirac CP phase vary within the $3\sigma$ level~\cite{Esteban:2020cvm}. The results from this scan concerning the magnitude of the GNI interactions and the reach of LHC searches do not differ from those obtained in the scenario with CP conservation. Moreover, the LFV observables continue exhibiting a strong correlation as the one displayed in Fig. \ref{fig:LFV}.

\section{Summary}\label{sec:conclusions}
The simplest model leading to Dirac neutrino masses at one loop level is obtained by adding a pair of charged scalars to the SM along with three right-handed neutrinos.
In this work we have assumed that the Diracness of the neutrinos is protected by only one extra $U(1)_{B-L}$ gauge symmetry, with the anomalies canceled by the SM leptons and the three right-handed neutrinos. 
We studied the interesting phenomenological features of the model such as the neutrino mass generation, charged LFV processes and signatures at the LHC associated to the charged scalars and the new $B-L$ gauge boson. 
Furthermore, we reported the expressions for the general neutrino-electron interactions and identified the regions of parameter space that may be explored in future solar neutrino experiments.
Remarkably, future searches for charged scalars at the LHC will probe the entire parameter space considered in this analysis.

\section*{Acknowledgments}
The work of DR and OZ is partially supported by Sostenibilidad-UdeA and UdeA/CODI Grants 2017-16286 and 2020-33177.

\bibliographystyle{apsrev4-1long}
\bibliography{references}

\begin{thebibliography}{10}%
\makeatletter
\providecommand \@ifxundefined [1]{%
 \ifx #1\undefined \expandafter \@firstoftwo
 \else \expandafter \@secondoftwo
\fi
}%
\providecommand \@ifnum [1]{%
 \ifnum #1\expandafter \@firstoftwo
 \else \expandafter \@secondoftwo
\fi
}%
\providecommand \enquote [1]{``#1''}%
\providecommand \bibnamefont  [1]{#1}%
\providecommand \bibfnamefont [1]{#1}%
\providecommand \citenamefont [1]{#1}%
\providecommand\href[0]{\@sanitize\@href}%
\providecommand\@href[1]{\endgroup\@@startlink{#1}\endgroup\@@href}%
\providecommand\@@href[1]{#1\@@endlink}%
\providecommand \@sanitize [0]{\begingroup\catcode`\&12\catcode`\#12\relax}%
\@ifxundefined \pdfoutput {\@firstoftwo}{%
 \@ifnum{\z@=\pdfoutput}{\@firstoftwo}{\@secondoftwo}%
}{%
 \providecommand\@@startlink[1]{\leavevmode\special{html:<a href="#1">}}%
 \providecommand\@@endlink[0]{\special{html:</a>}}%
}{%
 \providecommand\@@startlink[1]{%
  \leavevmode
  \pdfstartlink
   attr{/Border[0 0 1 ]/H/I/C[0 1 1]}%
   user{/Subtype/Link/A<</Type/Action/S/URI/URI(#1)>>}%
  \relax
 }%
 \providecommand\@@endlink[0]{\pdfendlink}%
}%
\providecommand \url  [0]{\begingroup\@sanitize \@url }%
\providecommand \@url [1]{\endgroup\@href {#1}{\urlprefix}}%
\providecommand \urlprefix [0]{URL }%
\providecommand \Eprint[0]{\href }%
\@ifxundefined \urlstyle {%
  \providecommand \doi [1]{doi:\discretionary{}{}{}#1}%
}{%
  \providecommand \doi [0]{doi:\discretionary{}{}{}\begingroup
  \urlstyle{rm}\Url }%
}%
\providecommand \doibase [0]{http://dx.doi.org/}%
\providecommand \Doi[1]{\href{\doibase#1}}%
\providecommand \bibAnnote [3]{%
  \BibitemShut{#1}%
  \begin{quotation}\noindent
    \textsc{Key:}\ #2\\\textsc{Annotation:}\ #3%
  \end{quotation}%
}%
\providecommand \bibAnnoteFile [2]{%
  \IfFileExists{#2}{\bibAnnote {#1} {#2} {\input{#2}}}{}%
}%
\providecommand \typeout [0]{\immediate \write \m@ne }%
\providecommand \selectlanguage [0]{\@gobble}%
\providecommand \bibinfo [0]{\@secondoftwo}%
\providecommand \bibfield [0]{\@secondoftwo}%
\providecommand \translation [1]{[#1]}%
\providecommand \BibitemOpen[0]{}%
\providecommand \bibitemStop [0]{}%
\providecommand \bibitemNoStop [0]{.\EOS\space}%
\providecommand \EOS [0]{\spacefactor3000\relax}%
\providecommand \BibitemShut [1]{\csname bibitem#1\endcsname}%
\bibitem{Fukuda:1998mi}%
  \BibitemOpen
  \bibfield{author}{%
  \bibinfo {author} {\bibfnamefont{Y.}~\bibnamefont{Fukuda}} \emph{et~al.}
  (\bibinfo {collaboration} {Super-Kamiokande Collaboration}),\ }%
  \bibfield{title}{%
  \enquote{\bibinfo {title} {{Evidence for oscillation of atmospheric
  neutrinos}},}\ }%
  \bibfield{journal}{%
  \Doi{10.1103/PhysRevLett.81.1562}{\bibinfo {journal} {Phys.Rev.Lett.}}\ }%
  \textbf{\bibinfo {volume} {81}},\ \bibinfo {pages} {1562--1567} (\bibinfo
  {year} {1998}),\
  \Eprint{http://arxiv.org/abs/hep-ex/9807003}{arXiv:hep-ex/9807003 [hep-ex]}%
  \bibAnnoteFile{NoStop}{Fukuda:1998mi}%
\bibitem{Ahmad:2002jz}%
  \BibitemOpen
  \bibfield{author}{%
  \bibinfo {author} {\bibfnamefont{Q.R.}\ \bibnamefont{Ahmad}} \emph{et~al.}
  (\bibinfo {collaboration} {SNO Collaboration}),\ }%
  \bibfield{title}{%
  \enquote{\bibinfo {title} {{Direct evidence for neutrino flavor
  transformation from neutral current interactions in the Sudbury Neutrino
  Observatory}},}\ }%
  \bibfield{journal}{%
  \Doi{10.1103/PhysRevLett.89.011301}{\bibinfo {journal} {Phys.Rev.Lett.}}\ }%
  \textbf{\bibinfo {volume} {89}},\ \bibinfo {pages} {011301} (\bibinfo {year}
  {2002}),\ \Eprint{http://arxiv.org/abs/nucl-ex/0204008}{arXiv:nucl-ex/0204008
  [nucl-ex]}%
  \bibAnnoteFile{NoStop}{Ahmad:2002jz}%
\bibitem{Dolinski:2019nrj}%
  \BibitemOpen
  \bibfield{author}{%
  \bibinfo {author} {\bibfnamefont{Michelle~J.}\ \bibnamefont{Dolinski}},
  \bibinfo {author} {\bibfnamefont{Alan W.~P.}\ \bibnamefont{Poon}},\ and\
  \bibinfo {author} {\bibfnamefont{Werner}\ \bibnamefont{Rodejohann}},\ }%
  \bibfield{title}{%
  \enquote{\bibinfo {title} {{Neutrinoless Double-Beta Decay: Status and
  Prospects}},}\ }%
  \bibfield{journal}{%
  \Doi{10.1146/annurev-nucl-101918-023407}{\bibinfo {journal} {Ann. Rev. Nucl.
  Part. Sci.}}\ }%
  \textbf{\bibinfo {volume} {69}},\ \bibinfo {pages} {219--251} (\bibinfo
  {year} {2019}),\ \Eprint{http://arxiv.org/abs/1902.04097}{arXiv:1902.04097
  [nucl-ex]}%
  \bibAnnoteFile{NoStop}{Dolinski:2019nrj}%
\bibitem{Lindner:2016bgg}%
  \BibitemOpen
  \bibfield{author}{%
  \bibinfo {author} {\bibfnamefont{Manfred}\ \bibnamefont{Lindner}}, \bibinfo
  {author} {\bibfnamefont{Moritz}\ \bibnamefont{Platscher}},\ and\ \bibinfo
  {author} {\bibfnamefont{Farinaldo~S.}\ \bibnamefont{Queiroz}},\ }%
  \bibfield{title}{%
  \enquote{\bibinfo {title} {{A Call for New Physics : The Muon Anomalous
  Magnetic Moment and Lepton Flavor Violation}},}\ }%
  \bibfield{journal}{%
  \Doi{10.1016/j.physrep.2017.12.001}{\bibinfo {journal} {Phys. Rept.}}\ }%
  \textbf{\bibinfo {volume} {731}},\ \bibinfo {pages} {1--82} (\bibinfo {year}
  {2018}),\ \Eprint{http://arxiv.org/abs/1610.06587}{arXiv:1610.06587
  [hep-ph]}%
  \bibAnnoteFile{NoStop}{Lindner:2016bgg}%
\bibitem{Calibbi:2017uvl}%
  \BibitemOpen
  \bibfield{author}{%
  \bibinfo {author} {\bibfnamefont{Lorenzo}\ \bibnamefont{Calibbi}}\ and\
  \bibinfo {author} {\bibfnamefont{Giovanni}\ \bibnamefont{Signorelli}},\ }%
  \bibfield{title}{%
  \enquote{\bibinfo {title} {{Charged Lepton Flavour Violation: An Experimental
  and Theoretical Introduction}},}\ }%
  \bibfield{journal}{%
  \Doi{10.1393/ncr/i2018-10144-0}{\bibinfo {journal} {Riv. Nuovo Cim.}}\ }%
  \textbf{\bibinfo {volume} {41}},\ \bibinfo {pages} {1} (\bibinfo {year}
  {2018}),\ \Eprint{http://arxiv.org/abs/1709.00294}{arXiv:1709.00294
  [hep-ph]}%
  \bibAnnoteFile{NoStop}{Calibbi:2017uvl}%
\bibitem{Djurcic:2015vqa}%
  \BibitemOpen
  \bibfield{author}{%
  \bibinfo {author} {\bibfnamefont{Zelimir}\ \bibnamefont{Djurcic}}
  \emph{et~al.} (\bibinfo {collaboration} {JUNO}),\ }%
  \bibfield{title}{%
  \enquote{\bibinfo {title} {{JUNO Conceptual Design Report}},}\ }%
   (\bibinfo {month} {8}\ \bibinfo {year} {2015}),\
  \Eprint{http://arxiv.org/abs/1508.07166}{arXiv:1508.07166 [physics.ins-det]}%
  \bibAnnoteFile{NoStop}{Djurcic:2015vqa}%
\bibitem{Acciarri:2015uup}%
  \BibitemOpen
  \bibfield{author}{%
  \bibinfo {author} {\bibfnamefont{R.}~\bibnamefont{Acciarri}} \emph{et~al.}
  (\bibinfo {collaboration} {DUNE}),\ }%
  \bibfield{title}{%
  \enquote{\bibinfo {title} {{Long-Baseline Neutrino Facility (LBNF) and Deep
  Underground Neutrino Experiment (DUNE)}: {Conceptual Design Report, Volume 2:
  The Physics Program for DUNE at LBNF}},}\ }%
   (\bibinfo {month} {12}\ \bibinfo {year} {2015}),\
  \Eprint{http://arxiv.org/abs/1512.06148}{arXiv:1512.06148 [physics.ins-det]}%
  \bibAnnoteFile{NoStop}{Acciarri:2015uup}%
\bibitem{Abe:2018uyc}%
  \BibitemOpen
  \bibfield{author}{%
  \bibinfo {author} {\bibfnamefont{K.}~\bibnamefont{Abe}} \emph{et~al.}
  (\bibinfo {collaboration} {Hyper-Kamiokande}),\ }%
  \bibfield{title}{%
  \enquote{\bibinfo {title} {{Hyper-Kamiokande Design Report}},}\ }%
   (\bibinfo {month} {5}\ \bibinfo {year} {2018}),\
  \Eprint{http://arxiv.org/abs/1805.04163}{arXiv:1805.04163 [physics.ins-det]}%
  \bibAnnoteFile{NoStop}{Abe:2018uyc}%
\bibitem{Baldini:2013ke}%
  \BibitemOpen
  \bibfield{author}{%
  \bibinfo {author} {\bibfnamefont{A.~M.}\ \bibnamefont{Baldini}}
  \emph{et~al.},\ }%
  \bibfield{title}{%
  \enquote{\bibinfo {title} {{MEG Upgrade Proposal}},}\ }%
   (\bibinfo {month} {1}\ \bibinfo {year} {2013}),\
  \Eprint{http://arxiv.org/abs/1301.7225}{arXiv:1301.7225 [physics.ins-det]}%
  \bibAnnoteFile{NoStop}{Baldini:2013ke}%
\bibitem{Aushev:2010bq}%
  \BibitemOpen
  \bibfield{author}{%
  \bibinfo {author} {\bibfnamefont{T.}~\bibnamefont{Aushev}} \emph{et~al.},\ }%
  \bibfield{title}{%
  \enquote{\bibinfo {title} {{Physics at Super B Factory}},}\ }%
   (\bibinfo {month} {2}\ \bibinfo {year} {2010}),\
  \Eprint{http://arxiv.org/abs/1002.5012}{arXiv:1002.5012 [hep-ex]}%
  \bibAnnoteFile{NoStop}{Aushev:2010bq}%
\bibitem{Abrams:2012er}%
  \BibitemOpen
  \bibfield{author}{%
  \bibinfo {author} {\bibfnamefont{R.~J.}\ \bibnamefont{Abrams}} \emph{et~al.}
  (\bibinfo {collaboration} {Mu2e}),\ }%
  \bibfield{title}{%
  \enquote{\bibinfo {title} {{Mu2e Conceptual Design Report}},}\ }%
   (\bibinfo {month} {11}\ \bibinfo {year} {2012}),\
  \Eprint{http://arxiv.org/abs/1211.7019}{arXiv:1211.7019 [physics.ins-det]}%
  \bibAnnoteFile{NoStop}{Abrams:2012er}%
\bibitem{Das:2017nvm}%
  \BibitemOpen
  \bibfield{author}{%
  \bibinfo {author} {\bibfnamefont{Arindam}\ \bibnamefont{Das}}\ and\ \bibinfo
  {author} {\bibfnamefont{Nobuchika}\ \bibnamefont{Okada}},\ }%
  \bibfield{title}{%
  \enquote{\bibinfo {title} {{Bounds on heavy Majorana neutrinos in type-I
  seesaw and implications for collider searches}},}\ }%
  \bibfield{journal}{%
  \Doi{10.1016/j.physletb.2017.09.042}{\bibinfo {journal} {Phys. Lett. B}}\ }%
  \textbf{\bibinfo {volume} {774}},\ \bibinfo {pages} {32--40} (\bibinfo {year}
  {2017}),\ \Eprint{http://arxiv.org/abs/1702.04668}{arXiv:1702.04668
  [hep-ph]}%
  \bibAnnoteFile{NoStop}{Das:2017nvm}%
\bibitem{Das:2019fee}%
  \BibitemOpen
  \bibfield{author}{%
  \bibinfo {author} {\bibfnamefont{Arindam}\ \bibnamefont{Das}}, \bibinfo
  {author} {\bibfnamefont{P.~S.~Bhupal}\ \bibnamefont{Dev}},\ and\ \bibinfo
  {author} {\bibfnamefont{Nobuchika}\ \bibnamefont{Okada}},\ }%
  \bibfield{title}{%
  \enquote{\bibinfo {title} {{Long-lived TeV-scale right-handed neutrino
  production at the LHC in gauged $U(1)_X$ model}},}\ }%
  \bibfield{journal}{%
  \Doi{10.1016/j.physletb.2019.135052}{\bibinfo {journal} {Phys. Lett. B}}\ }%
  \textbf{\bibinfo {volume} {799}},\ \bibinfo {pages} {135052} (\bibinfo {year}
  {2019}),\ \Eprint{http://arxiv.org/abs/1906.04132}{arXiv:1906.04132
  [hep-ph]}%
  \bibAnnoteFile{NoStop}{Das:2019fee}%
\bibitem{Das:2012ze}%
  \BibitemOpen
  \bibfield{author}{%
  \bibinfo {author} {\bibfnamefont{Arindam}\ \bibnamefont{Das}}\ and\ \bibinfo
  {author} {\bibfnamefont{Nobuchika}\ \bibnamefont{Okada}},\ }%
  \bibfield{title}{%
  \enquote{\bibinfo {title} {{Inverse seesaw neutrino signatures at the LHC and
  ILC}},}\ }%
  \bibfield{journal}{%
  \Doi{10.1103/PhysRevD.88.113001}{\bibinfo {journal} {Phys. Rev. D}}\ }%
  \textbf{\bibinfo {volume} {88}},\ \bibinfo {pages} {113001} (\bibinfo {year}
  {2013}),\ \Eprint{http://arxiv.org/abs/1207.3734}{arXiv:1207.3734 [hep-ph]}%
  \bibAnnoteFile{NoStop}{Das:2012ze}%
\bibitem{Minkowski:1977sc}%
  \BibitemOpen
  \bibfield{author}{%
  \bibinfo {author} {\bibfnamefont{Peter}\ \bibnamefont{Minkowski}},\ }%
  \bibfield{title}{%
  \enquote{\bibinfo {title} {{$\mu \to e\gamma$ at a Rate of One Out of
  $10^{9}$ Muon Decays?}}.}\ }%
  \bibfield{journal}{%
  \Doi{10.1016/0370-2693(77)90435-X}{\bibinfo {journal} {Phys. Lett.}}\ }%
  \textbf{\bibinfo {volume} {67B}},\ \bibinfo {pages} {421--428} (\bibinfo
  {year} {1977})%
  \bibAnnoteFile{NoStop}{Minkowski:1977sc}%
\bibitem{Yanagida:1979as}%
  \BibitemOpen
  \bibfield{author}{%
  \bibinfo {author} {\bibfnamefont{Tsutomu}\ \bibnamefont{Yanagida}},\ }%
  \bibfield{title}{%
  \enquote{\bibinfo {title} {{Horizontal gauge symmetry and masses of
  neutrinos}},}\ }%
  \bibfield{booktitle}{%
  \emph{\bibinfo {booktitle} {{Proceedings: Workshop on the Unified Theories
  and the Baryon Number in the Universe: Tsukuba, Japan, February 13-14,
  1979}}},\ }%
  \bibfield{journal}{%
  \bibinfo {journal} {Conf. Proc.}\ }%
  \textbf{\bibinfo {volume} {C7902131}},\ \bibinfo {pages} {95--99} (\bibinfo
  {year} {1979})%
  \bibAnnoteFile{NoStop}{Yanagida:1979as}%
\bibitem{GellMann:1980vs}%
  \BibitemOpen
  \bibfield{author}{%
  \bibinfo {author} {\bibfnamefont{Murray}\ \bibnamefont{Gell-Mann}}, \bibinfo
  {author} {\bibfnamefont{Pierre}\ \bibnamefont{Ramond}},\ and\ \bibinfo
  {author} {\bibfnamefont{Richard}\ \bibnamefont{Slansky}},\ }%
  \bibfield{title}{%
  \enquote{\bibinfo {title} {{Complex Spinors and Unified Theories}},}\ }%
  \bibfield{booktitle}{%
  \emph{\bibinfo {booktitle} {{Supergravity Workshop Stony Brook, New York,
  September 27-28, 1979}}},\ }%
  \bibfield{journal}{%
  \bibinfo {journal} {Conf. Proc.}\ }%
  \textbf{\bibinfo {volume} {C790927}},\ \bibinfo {pages} {315--321} (\bibinfo
  {year} {1979}),\ \Eprint{http://arxiv.org/abs/1306.4669}{arXiv:1306.4669
  [hep-th]}%
  \bibAnnoteFile{NoStop}{GellMann:1980vs}%
\bibitem{Mohapatra:1979ia}%
  \BibitemOpen
  \bibfield{author}{%
  \bibinfo {author} {\bibfnamefont{Rabindra~N.}\ \bibnamefont{Mohapatra}}\ and\
  \bibinfo {author} {\bibfnamefont{Goran}\ \bibnamefont{Senjanovic}},\ }%
  \bibfield{title}{%
  \enquote{\bibinfo {title} {{Neutrino Mass and Spontaneous Parity
  Nonconservation}},}\ }%
  \bibfield{journal}{%
  \Doi{10.1103/PhysRevLett.44.912}{\bibinfo {journal} {Phys. Rev. Lett.}}\ }%
  \textbf{\bibinfo {volume} {44}},\ \bibinfo {pages} {912} (\bibinfo {year}
  {1980}),\ \bibinfo {note} {[,231(1979)]}%
  \bibAnnoteFile{NoStop}{Mohapatra:1979ia}%
\bibitem{Davidson:1978pm}%
  \BibitemOpen
  \bibfield{author}{%
  \bibinfo {author} {\bibfnamefont{Aharon}\ \bibnamefont{Davidson}},\ }%
  \bibfield{title}{%
  \enquote{\bibinfo {title} {{$B-L$ as the fourth color within an
  $\mathrm{SU}(2)_L \times \mathrm{U}(1)_R \times \mathrm{U}(1)$ model}},}\ }%
  \bibfield{journal}{%
  \Doi{10.1103/PhysRevD.20.776}{\bibinfo {journal} {Phys. Rev. D}}\ }%
  \textbf{\bibinfo {volume} {20}},\ \bibinfo {pages} {776} (\bibinfo {year}
  {1979})%
  \bibAnnoteFile{NoStop}{Davidson:1978pm}%
\bibitem{Calle:2018ovc}%
  \BibitemOpen
  \bibfield{author}{%
  \bibinfo {author} {\bibfnamefont{Julian}\ \bibnamefont{Calle}}, \bibinfo
  {author} {\bibfnamefont{Diego}\ \bibnamefont{Restrepo}}, \bibinfo {author}
  {\bibfnamefont{Carlos~E.}\ \bibnamefont{Yaguna}},\ and\ \bibinfo {author}
  {\bibfnamefont{Óscar}\ \bibnamefont{Zapata}},\ }%
  \bibfield{title}{%
  \enquote{\bibinfo {title} {{Minimal radiative Dirac neutrino mass models}},}\
  }%
  \bibfield{journal}{%
  \Doi{10.1103/PhysRevD.99.075008}{\bibinfo {journal} {Phys. Rev. D}}\ }%
  \textbf{\bibinfo {volume} {99}},\ \bibinfo {pages} {075008} (\bibinfo {year}
  {2019}),\ \Eprint{http://arxiv.org/abs/1812.05523}{arXiv:1812.05523
  [hep-ph]}%
  \bibAnnoteFile{NoStop}{Calle:2018ovc}%
\bibitem{Bonilla:2018ynb}%
  \BibitemOpen
  \bibfield{author}{%
  \bibinfo {author} {\bibfnamefont{Cesar}\ \bibnamefont{Bonilla}}, \bibinfo
  {author} {\bibfnamefont{Salvador}\ \bibnamefont{Centelles-Chuliá}}, \bibinfo
  {author} {\bibfnamefont{Ricardo}\ \bibnamefont{Cepedello}}, \bibinfo {author}
  {\bibfnamefont{Eduardo}\ \bibnamefont{Peinado}},\ and\ \bibinfo {author}
  {\bibfnamefont{Rahul}\ \bibnamefont{Srivastava}},\ }%
  \bibfield{title}{%
  \enquote{\bibinfo {title} {{Dark matter stability and Dirac neutrinos using
  only Standard Model symmetries}},}\ }%
  \bibfield{journal}{%
  \Doi{10.1103/PhysRevD.101.033011}{\bibinfo {journal} {Phys. Rev. D}}\ }%
  \textbf{\bibinfo {volume} {101}},\ \bibinfo {pages} {033011} (\bibinfo {year}
  {2020}),\ \Eprint{http://arxiv.org/abs/1812.01599}{arXiv:1812.01599
  [hep-ph]}%
  \bibAnnoteFile{NoStop}{Bonilla:2018ynb}%
\bibitem{Saad:2019bqf}%
  \BibitemOpen
  \bibfield{author}{%
  \bibinfo {author} {\bibfnamefont{Shaikh}\ \bibnamefont{Saad}},\ }%
  \bibfield{title}{%
  \enquote{\bibinfo {title} {{Simplest Radiative Dirac Neutrino Mass
  Models}},}\ }%
  \bibfield{journal}{%
  \Doi{10.1016/j.nuclphysb.2019.114636}{\bibinfo {journal} {Nucl. Phys. B}}\ }%
  \textbf{\bibinfo {volume} {943}},\ \bibinfo {pages} {114636} (\bibinfo {year}
  {2019}),\ \Eprint{http://arxiv.org/abs/1902.07259}{arXiv:1902.07259
  [hep-ph]}%
  \bibAnnoteFile{NoStop}{Saad:2019bqf}%
\bibitem{Cai:2017jrq}%
  \BibitemOpen
  \bibfield{author}{%
  \bibinfo {author} {\bibfnamefont{Yi}~\bibnamefont{Cai}}, \bibinfo {author}
  {\bibfnamefont{Juan}\ \bibnamefont{Herrero-Garc\'\i{}a}}, \bibinfo {author}
  {\bibfnamefont{Michael~A.}\ \bibnamefont{Schmidt}}, \bibinfo {author}
  {\bibfnamefont{Avelino}\ \bibnamefont{Vicente}},\ and\ \bibinfo {author}
  {\bibfnamefont{Raymond~R.}\ \bibnamefont{Volkas}},\ }%
  \bibfield{title}{%
  \enquote{\bibinfo {title} {{From the trees to the forest: a review of
  radiative neutrino mass models}},}\ }%
  \bibfield{journal}{%
  \Doi{10.3389/fphy.2017.00063}{\bibinfo {journal} {Front. in Phys.}}\ }%
  \textbf{\bibinfo {volume} {5}},\ \bibinfo {pages} {63} (\bibinfo {year}
  {2017}),\ \Eprint{http://arxiv.org/abs/1706.08524}{arXiv:1706.08524
  [hep-ph]}%
  \bibAnnoteFile{NoStop}{Cai:2017jrq}%
\bibitem{Nasri:2001ax}%
  \BibitemOpen
  \bibfield{author}{%
  \bibinfo {author} {\bibfnamefont{Salah}\ \bibnamefont{Nasri}}\ and\ \bibinfo
  {author} {\bibfnamefont{Sherif}\ \bibnamefont{Moussa}},\ }%
  \bibfield{title}{%
  \enquote{\bibinfo {title} {{Model for small neutrino masses at the TeV
  scale}},}\ }%
  \bibfield{journal}{%
  \Doi{10.1142/S0217732302007119}{\bibinfo {journal} {Mod. Phys. Lett. A}}\ }%
  \textbf{\bibinfo {volume} {17}},\ \bibinfo {pages} {771--778} (\bibinfo
  {year} {2002}),\
  \Eprint{http://arxiv.org/abs/hep-ph/0106107}{arXiv:hep-ph/0106107}%
  \bibAnnoteFile{NoStop}{Nasri:2001ax}%
\bibitem{Kanemura:2011jj}%
  \BibitemOpen
  \bibfield{author}{%
  \bibinfo {author} {\bibfnamefont{Shinya}\ \bibnamefont{Kanemura}}, \bibinfo
  {author} {\bibfnamefont{Takehiro}\ \bibnamefont{Nabeshima}},\ and\ \bibinfo
  {author} {\bibfnamefont{Hiroaki}\ \bibnamefont{Sugiyama}},\ }%
  \bibfield{title}{%
  \enquote{\bibinfo {title} {{Neutrino Masses from Loop-Induced Dirac Yukawa
  Couplings}},}\ }%
  \bibfield{journal}{%
  \Doi{10.1016/j.physletb.2011.07.047}{\bibinfo {journal} {Phys. Lett.}}\ }%
  \textbf{\bibinfo {volume} {B703}},\ \bibinfo {pages} {66--70} (\bibinfo
  {year} {2011}),\ \Eprint{http://arxiv.org/abs/1106.2480}{arXiv:1106.2480
  [hep-ph]}%
  \bibAnnoteFile{NoStop}{Kanemura:2011jj}%
\bibitem{Jana:2019mez}%
  \BibitemOpen
  \bibfield{author}{%
  \bibinfo {author} {\bibfnamefont{Sudip}\ \bibnamefont{Jana}}, \bibinfo
  {author} {\bibfnamefont{P.~K.}\ \bibnamefont{Vishnu}},\ and\ \bibinfo
  {author} {\bibfnamefont{Shaikh}\ \bibnamefont{Saad}},\ }%
  \bibfield{title}{%
  \enquote{\bibinfo {title} {{Minimal dirac neutrino mass models from $\hbox
  {U}(1)_{\mathrm{R}}$ gauge symmetry and left\textendash{}right asymmetry at
  colliders}},}\ }%
  \bibfield{journal}{%
  \Doi{10.1140/epjc/s10052-019-7441-9}{\bibinfo {journal} {Eur. Phys. J. C}}\
  }%
  \textbf{\bibinfo {volume} {79}},\ \bibinfo {pages} {916} (\bibinfo {year}
  {2019}),\ \Eprint{http://arxiv.org/abs/1904.07407}{arXiv:1904.07407
  [hep-ph]}%
  \bibAnnoteFile{NoStop}{Jana:2019mez}%
\bibitem{Yao:2018ekp}%
  \BibitemOpen
  \bibfield{author}{%
  \bibinfo {author} {\bibfnamefont{Chang-Yuan}\ \bibnamefont{Yao}}\ and\
  \bibinfo {author} {\bibfnamefont{Gui-Jun}\ \bibnamefont{Ding}},\ }%
  \bibfield{title}{%
  \enquote{\bibinfo {title} {{Systematic Analysis of Dirac Neutrino Masses at
  Dimension Five}},}\ }%
   (\bibinfo {year} {2018}),\
  \Eprint{http://arxiv.org/abs/1802.05231}{arXiv:1802.05231 [hep-ph]}%
  \bibAnnoteFile{NoStop}{Yao:2018ekp}%
\bibitem{Jana:2019mgj}%
  \BibitemOpen
  \bibfield{author}{%
  \bibinfo {author} {\bibfnamefont{Sudip}\ \bibnamefont{Jana}}, \bibinfo
  {author} {\bibfnamefont{P.K.}\ \bibnamefont{Vishnu}},\ and\ \bibinfo {author}
  {\bibfnamefont{Shaikh}\ \bibnamefont{Saad}},\ }%
  \bibfield{title}{%
  \enquote{\bibinfo {title} {{Minimal Realizations of Dirac Neutrino Mass from
  Generic One-loop and Two-loop Topologies at $d=5$}},}\ }%
  \bibfield{journal}{%
  \Doi{10.1088/1475-7516/2020/04/018}{\bibinfo {journal} {JCAP}}\ }%
  \textbf{\bibinfo {volume} {04}},\ \bibinfo {pages} {018} (\bibinfo {year}
  {2020}),\ \Eprint{http://arxiv.org/abs/1910.09537}{arXiv:1910.09537
  [hep-ph]}%
  \bibAnnoteFile{NoStop}{Jana:2019mgj}%
\bibitem{Zee:1980ai}%
  \BibitemOpen
  \bibfield{author}{%
  \bibinfo {author} {\bibfnamefont{A.}~\bibnamefont{Zee}},\ }%
  \bibfield{title}{%
  \enquote{\bibinfo {title} {{A Theory of Lepton Number Violation, Neutrino
  Majorana Mass, and Oscillation}},}\ }%
  \bibfield{journal}{%
  \Doi{10.1016/0370-2693(80)90349-4}{\bibinfo {journal} {Phys. Lett. B}}\ }%
  \textbf{\bibinfo {volume} {93}},\ \bibinfo {pages} {389} (\bibinfo {year}
  {1980}),\ \bibinfo {note} {[Erratum: Phys.Lett.B 95, 461 (1980)]}%
  \bibAnnoteFile{NoStop}{Zee:1980ai}%
\bibitem{Petcov:1982en}%
  \BibitemOpen
  \bibfield{author}{%
  \bibinfo {author} {\bibfnamefont{S.~T.}\ \bibnamefont{Petcov}},\ }%
  \bibfield{title}{%
  \enquote{\bibinfo {title} {{Remarks on the Zee Model of Neutrino Mixing (mu
  ---\ensuremath{>} e gamma, Heavy Neutrino ---\ensuremath{>} Light Neutrino
  gamma, etc.)}},}\ }%
  \bibfield{journal}{%
  \Doi{10.1016/0370-2693(82)90526-3}{\bibinfo {journal} {Phys. Lett. B}}\ }%
  \textbf{\bibinfo {volume} {115}},\ \bibinfo {pages} {401--406} (\bibinfo
  {year} {1982})%
  \bibAnnoteFile{NoStop}{Petcov:1982en}%
\bibitem{Bergmann:1999rz}%
  \BibitemOpen
  \bibfield{author}{%
  \bibinfo {author} {\bibfnamefont{Sven}\ \bibnamefont{Bergmann}}, \bibinfo
  {author} {\bibfnamefont{Yuval}\ \bibnamefont{Grossman}},\ and\ \bibinfo
  {author} {\bibfnamefont{Enrico}\ \bibnamefont{Nardi}},\ }%
  \bibfield{title}{%
  \enquote{\bibinfo {title} {{Neutrino propagation in matter with general
  interactions}},}\ }%
  \bibfield{journal}{%
  \Doi{10.1103/PhysRevD.60.093008}{\bibinfo {journal} {Phys. Rev. D}}\ }%
  \textbf{\bibinfo {volume} {60}},\ \bibinfo {pages} {093008} (\bibinfo {year}
  {1999}),\ \Eprint{http://arxiv.org/abs/hep-ph/9903517}{arXiv:hep-ph/9903517}%
  \bibAnnoteFile{NoStop}{Bergmann:1999rz}%
\bibitem{Khan:2019jvr}%
  \BibitemOpen
  \bibfield{author}{%
  \bibinfo {author} {\bibfnamefont{Amir~N.}\ \bibnamefont{Khan}}, \bibinfo
  {author} {\bibfnamefont{Werner}\ \bibnamefont{Rodejohann}},\ and\ \bibinfo
  {author} {\bibfnamefont{Xun-Jie}\ \bibnamefont{Xu}},\ }%
  \bibfield{title}{%
  \enquote{\bibinfo {title} {{Borexino and General Neutrino Interactions}},}\
  }%
   (\bibinfo {year} {2019}),\
  \Eprint{http://arxiv.org/abs/1906.12102}{arXiv:1906.12102 [hep-ph]}%
  \bibAnnoteFile{NoStop}{Khan:2019jvr}%
\bibitem{Bischer:2019ttk}%
  \BibitemOpen
  \bibfield{author}{%
  \bibinfo {author} {\bibfnamefont{Ingolf}\ \bibnamefont{Bischer}}\ and\
  \bibinfo {author} {\bibfnamefont{Werner}\ \bibnamefont{Rodejohann}},\ }%
  \bibfield{title}{%
  \enquote{\bibinfo {title} {{General neutrino interactions from an effective
  field theory perspective}},}\ }%
  \bibfield{journal}{%
  \Doi{10.1016/j.nuclphysb.2019.114746}{\bibinfo {journal} {Nucl. Phys. B}}\ }%
  \textbf{\bibinfo {volume} {947}},\ \bibinfo {pages} {114746} (\bibinfo {year}
  {2019}),\ \Eprint{http://arxiv.org/abs/1905.08699}{arXiv:1905.08699
  [hep-ph]}%
  \bibAnnoteFile{NoStop}{Bischer:2019ttk}%
\bibitem{Jinno:2016knw}%
  \BibitemOpen
  \bibfield{author}{%
  \bibinfo {author} {\bibfnamefont{Ryusuke}\ \bibnamefont{Jinno}}\ and\
  \bibinfo {author} {\bibfnamefont{Masahiro}\ \bibnamefont{Takimoto}},\ }%
  \bibfield{title}{%
  \enquote{\bibinfo {title} {{Probing a classically conformal B-L model with
  gravitational waves}},}\ }%
  \bibfield{journal}{%
  \Doi{10.1103/PhysRevD.95.015020}{\bibinfo {journal} {Phys. Rev. D}}\ }%
  \textbf{\bibinfo {volume} {95}},\ \bibinfo {pages} {015020} (\bibinfo {year}
  {2017}),\ \Eprint{http://arxiv.org/abs/1604.05035}{arXiv:1604.05035
  [hep-ph]}%
  \bibAnnoteFile{NoStop}{Jinno:2016knw}%
\bibitem{Chao:2017ilw}%
  \BibitemOpen
  \bibfield{author}{%
  \bibinfo {author} {\bibfnamefont{Wei}\ \bibnamefont{Chao}}, \bibinfo {author}
  {\bibfnamefont{Wen-Feng}\ \bibnamefont{Cui}}, \bibinfo {author}
  {\bibfnamefont{Huai-Ke}\ \bibnamefont{Guo}},\ and\ \bibinfo {author}
  {\bibfnamefont{Jing}\ \bibnamefont{Shu}},\ }%
  \bibfield{title}{%
  \enquote{\bibinfo {title} {{Gravitational wave imprint of new symmetry
  breaking}},}\ }%
  \bibfield{journal}{%
  \Doi{10.1088/1674-1137/abb4cb}{\bibinfo {journal} {Chin. Phys. C}}\ }%
  \textbf{\bibinfo {volume} {44}},\ \bibinfo {pages} {123102} (\bibinfo {year}
  {2020}),\ \Eprint{http://arxiv.org/abs/1707.09759}{arXiv:1707.09759
  [hep-ph]}%
  \bibAnnoteFile{NoStop}{Chao:2017ilw}%
\bibitem{Marzo:2018nov}%
  \BibitemOpen
  \bibfield{author}{%
  \bibinfo {author} {\bibfnamefont{Carlo}\ \bibnamefont{Marzo}}, \bibinfo
  {author} {\bibfnamefont{Luca}\ \bibnamefont{Marzola}},\ and\ \bibinfo
  {author} {\bibfnamefont{Ville}\ \bibnamefont{Vaskonen}},\ }%
  \bibfield{title}{%
  \enquote{\bibinfo {title} {{Phase transition and vacuum stability in the
  classically conformal B\textendash{}L model}},}\ }%
  \bibfield{journal}{%
  \Doi{10.1140/epjc/s10052-019-7076-x}{\bibinfo {journal} {Eur. Phys. J. C}}\
  }%
  \textbf{\bibinfo {volume} {79}},\ \bibinfo {pages} {601} (\bibinfo {year}
  {2019}),\ \Eprint{http://arxiv.org/abs/1811.11169}{arXiv:1811.11169
  [hep-ph]}%
  \bibAnnoteFile{NoStop}{Marzo:2018nov}%
\bibitem{Montero:2007cd}%
  \BibitemOpen
  \bibfield{author}{%
  \bibinfo {author} {\bibfnamefont{J.~C.}\ \bibnamefont{Montero}}\ and\
  \bibinfo {author} {\bibfnamefont{V.}~\bibnamefont{Pleitez}},\ }%
  \bibfield{title}{%
  \enquote{\bibinfo {title} {{Gauging U(1) symmetries and the number of
  right-handed neutrinos}},}\ }%
  \bibfield{journal}{%
  \Doi{10.1016/j.physletb.2009.03.065}{\bibinfo {journal} {Phys. Lett. B}}\ }%
  \textbf{\bibinfo {volume} {675}},\ \bibinfo {pages} {64--68} (\bibinfo {year}
  {2009}),\ \Eprint{http://arxiv.org/abs/0706.0473}{arXiv:0706.0473 [hep-ph]}%
  \bibAnnoteFile{NoStop}{Montero:2007cd}%
\bibitem{Ma:2014qra}%
  \BibitemOpen
  \bibfield{author}{%
  \bibinfo {author} {\bibfnamefont{Ernest}\ \bibnamefont{Ma}}\ and\ \bibinfo
  {author} {\bibfnamefont{Rahul}\ \bibnamefont{Srivastava}},\ }%
  \bibfield{title}{%
  \enquote{\bibinfo {title} {{Dirac or inverse seesaw neutrino masses with
  $B-L$ gauge symmetry and $S_3$ flavor symmetry}},}\ }%
  \bibfield{journal}{%
  \Doi{10.1016/j.physletb.2014.12.049}{\bibinfo {journal} {Phys. Lett.}}\ }%
  \textbf{\bibinfo {volume} {B741}},\ \bibinfo {pages} {217--222} (\bibinfo
  {year} {2015}),\ \Eprint{http://arxiv.org/abs/1411.5042}{arXiv:1411.5042
  [hep-ph]}%
  \bibAnnoteFile{NoStop}{Ma:2014qra}%
\bibitem{deSalas:2020pgw}%
  \BibitemOpen
  \bibfield{author}{%
  \bibinfo {author} {\bibfnamefont{P.~F.}\ \bibnamefont{de~Salas}}, \bibinfo
  {author} {\bibfnamefont{D.~V.}\ \bibnamefont{Forero}}, \bibinfo {author}
  {\bibfnamefont{S.}~\bibnamefont{Gariazzo}}, \bibinfo {author}
  {\bibfnamefont{P.}~\bibnamefont{Mart\'\i{}nez-Mirav\'e}}, \bibinfo {author}
  {\bibfnamefont{O.}~\bibnamefont{Mena}}, \bibinfo {author}
  {\bibfnamefont{C.~A.}\ \bibnamefont{Ternes}}, \bibinfo {author}
  {\bibfnamefont{M.}~\bibnamefont{T\'ortola}},\ and\ \bibinfo {author}
  {\bibfnamefont{J.~W.~F.}\ \bibnamefont{Valle}},\ }%
  \bibfield{title}{%
  \enquote{\bibinfo {title} {{2020 Global reassessment of the neutrino
  oscillation picture}},}\ }%
   (\bibinfo {month} {6}\ \bibinfo {year} {2020}),\
  \Eprint{http://arxiv.org/abs/2006.11237}{arXiv:2006.11237 [hep-ph]}%
  \bibAnnoteFile{NoStop}{deSalas:2020pgw}%
\bibitem{Esteban:2020cvm}%
  \BibitemOpen
  \bibfield{author}{%
  \bibinfo {author} {\bibfnamefont{Ivan}\ \bibnamefont{Esteban}}, \bibinfo
  {author} {\bibfnamefont{M.~C.}\ \bibnamefont{Gonzalez-Garcia}}, \bibinfo
  {author} {\bibfnamefont{Michele}\ \bibnamefont{Maltoni}}, \bibinfo {author}
  {\bibfnamefont{Thomas}\ \bibnamefont{Schwetz}},\ and\ \bibinfo {author}
  {\bibfnamefont{Albert}\ \bibnamefont{Zhou}},\ }%
  \bibfield{title}{%
  \enquote{\bibinfo {title} {{The fate of hints: updated global analysis of
  three-flavor neutrino oscillations}},}\ }%
  \bibfield{journal}{%
  \Doi{10.1007/JHEP09(2020)178}{\bibinfo {journal} {JHEP}}\ }%
  \textbf{\bibinfo {volume} {09}},\ \bibinfo {pages} {178} (\bibinfo {year}
  {2020}),\ \Eprint{http://arxiv.org/abs/2007.14792}{arXiv:2007.14792
  [hep-ph]}%
  \bibAnnoteFile{NoStop}{Esteban:2020cvm}%
\bibitem{Porod:2014xia}%
  \BibitemOpen
  \bibfield{author}{%
  \bibinfo {author} {\bibfnamefont{Werner}\ \bibnamefont{Porod}}, \bibinfo
  {author} {\bibfnamefont{Florian}\ \bibnamefont{Staub}},\ and\ \bibinfo
  {author} {\bibfnamefont{Avelino}\ \bibnamefont{Vicente}},\ }%
  \bibfield{title}{%
  \enquote{\bibinfo {title} {{A Flavor Kit for BSM models}},}\ }%
  \bibfield{journal}{%
  \Doi{10.1140/epjc/s10052-014-2992-2}{\bibinfo {journal} {Eur. Phys. J.}}\ }%
  \textbf{\bibinfo {volume} {C74}},\ \bibinfo {pages} {2992} (\bibinfo {year}
  {2014}),\ \Eprint{http://arxiv.org/abs/1405.1434}{arXiv:1405.1434 [hep-ph]}%
  \bibAnnoteFile{NoStop}{Porod:2014xia}%
\bibitem{TheMEG:2016wtm}%
  \BibitemOpen
  \bibfield{author}{%
  \bibinfo {author} {\bibfnamefont{A.~M.}\ \bibnamefont{Baldini}} \emph{et~al.}
  (\bibinfo {collaboration} {MEG}),\ }%
  \bibfield{title}{%
  \enquote{\bibinfo {title} {{Search for the lepton flavour violating decay
  $\mu ^+ \rightarrow \mathrm {e}^+ \gamma $ with the full dataset of the MEG
  experiment}},}\ }%
  \bibfield{journal}{%
  \Doi{10.1140/epjc/s10052-016-4271-x}{\bibinfo {journal} {Eur. Phys. J.}}\ }%
  \textbf{\bibinfo {volume} {C76}},\ \bibinfo {pages} {434} (\bibinfo {year}
  {2016}),\ \Eprint{http://arxiv.org/abs/1605.05081}{arXiv:1605.05081
  [hep-ex]}%
  \bibAnnoteFile{NoStop}{TheMEG:2016wtm}%
\bibitem{Aubert:2009ag}%
  \BibitemOpen
  \bibfield{author}{%
  \bibinfo {author} {\bibfnamefont{Bernard}\ \bibnamefont{Aubert}}
  \emph{et~al.} (\bibinfo {collaboration} {BaBar}),\ }%
  \bibfield{title}{%
  \enquote{\bibinfo {title} {{Searches for Lepton Flavor Violation in the
  Decays tau+- ---> e+- gamma and tau+- ---> mu+- gamma}},}\ }%
  \bibfield{journal}{%
  \Doi{10.1103/PhysRevLett.104.021802}{\bibinfo {journal} {Phys. Rev. Lett.}}\
  }%
  \textbf{\bibinfo {volume} {104}},\ \bibinfo {pages} {021802} (\bibinfo {year}
  {2010}),\ \Eprint{http://arxiv.org/abs/0908.2381}{arXiv:0908.2381 [hep-ex]}%
  \bibAnnoteFile{NoStop}{Aubert:2009ag}%
\bibitem{Bona:2007qt}%
  \BibitemOpen
  \bibfield{author}{%
  \bibinfo {author} {\bibfnamefont{M.}~\bibnamefont{Bona}} \emph{et~al.}
  (\bibinfo {collaboration} {SuperB}),\ }%
  \bibfield{title}{%
  \enquote{\bibinfo {title} {{SuperB: A High-Luminosity Asymmetric e+ e- Super
  Flavor Factory. Conceptual Design Report}},}\ }%
   (\bibinfo {month} {5}\ \bibinfo {year} {2007}),\
  \Eprint{http://arxiv.org/abs/0709.0451}{arXiv:0709.0451 [hep-ex]}%
  \bibAnnoteFile{NoStop}{Bona:2007qt}%
\bibitem{Miyazaki:2013yaa}%
  \BibitemOpen
  \bibfield{author}{%
  \bibinfo {author} {\bibfnamefont{Y.}~\bibnamefont{Miyazaki}} \emph{et~al.}
  (\bibinfo {collaboration} {Belle}),\ }%
  \bibfield{title}{%
  \enquote{\bibinfo {title} {{Search for Lepton-Flavor-Violating and
  Lepton-Number-Violating $\tau \to \ell h h^\prime$ Decay Modes}},}\ }%
  \bibfield{journal}{%
  \Doi{10.1016/j.physletb.2013.01.032}{\bibinfo {journal} {Phys. Lett. B}}\ }%
  \textbf{\bibinfo {volume} {719}},\ \bibinfo {pages} {346--353} (\bibinfo
  {year} {2013}),\ \Eprint{http://arxiv.org/abs/1206.5595}{arXiv:1206.5595
  [hep-ex]}%
  \bibAnnoteFile{NoStop}{Miyazaki:2013yaa}%
\bibitem{Bellgardt:1987du}%
  \BibitemOpen
  \bibfield{author}{%
  \bibinfo {author} {\bibfnamefont{U.}~\bibnamefont{Bellgardt}} \emph{et~al.}
  (\bibinfo {collaboration} {SINDRUM}),\ }%
  \bibfield{title}{%
  \enquote{\bibinfo {title} {{Search for the Decay mu+ ---> e+ e+ e-}},}\ }%
  \bibfield{journal}{%
  \Doi{10.1016/0550-3213(88)90462-2}{\bibinfo {journal} {Nucl. Phys. B}}\ }%
  \textbf{\bibinfo {volume} {299}},\ \bibinfo {pages} {1--6} (\bibinfo {year}
  {1988})%
  \bibAnnoteFile{NoStop}{Bellgardt:1987du}%
\bibitem{Hayasaka:2010np}%
  \BibitemOpen
  \bibfield{author}{%
  \bibinfo {author} {\bibfnamefont{K.}~\bibnamefont{Hayasaka}} \emph{et~al.},\
  }%
  \bibfield{title}{%
  \enquote{\bibinfo {title} {{Search for Lepton Flavor Violating Tau Decays
  into Three Leptons with 719 Million Produced Tau+Tau- Pairs}},}\ }%
  \bibfield{journal}{%
  \Doi{10.1016/j.physletb.2010.03.037}{\bibinfo {journal} {Phys. Lett. B}}\ }%
  \textbf{\bibinfo {volume} {687}},\ \bibinfo {pages} {139--143} (\bibinfo
  {year} {2010}),\ \Eprint{http://arxiv.org/abs/1001.3221}{arXiv:1001.3221
  [hep-ex]}%
  \bibAnnoteFile{NoStop}{Hayasaka:2010np}%
\bibitem{Dohmen:1993mp}%
  \BibitemOpen
  \bibfield{author}{%
  \bibinfo {author} {\bibfnamefont{C.}~\bibnamefont{Dohmen}} \emph{et~al.}
  (\bibinfo {collaboration} {SINDRUM II}),\ }%
  \bibfield{title}{%
  \enquote{\bibinfo {title} {{Test of lepton flavor conservation in mu
  ---\ensuremath{>} e conversion on titanium}},}\ }%
  \bibfield{journal}{%
  \Doi{10.1016/0370-2693(93)91383-X}{\bibinfo {journal} {Phys. Lett. B}}\ }%
  \textbf{\bibinfo {volume} {317}},\ \bibinfo {pages} {631--636} (\bibinfo
  {year} {1993})%
  \bibAnnoteFile{NoStop}{Dohmen:1993mp}%
\bibitem{Bertl:2006up}%
  \BibitemOpen
  \bibfield{author}{%
  \bibinfo {author} {\bibfnamefont{Wilhelm~H.}\ \bibnamefont{Bertl}}
  \emph{et~al.} (\bibinfo {collaboration} {SINDRUM II}),\ }%
  \bibfield{title}{%
  \enquote{\bibinfo {title} {{A Search for muon to electron conversion in
  muonic gold}},}\ }%
  \bibfield{journal}{%
  \Doi{10.1140/epjc/s2006-02582-x}{\bibinfo {journal} {Eur. Phys. J. C}}\ }%
  \textbf{\bibinfo {volume} {47}},\ \bibinfo {pages} {337--346} (\bibinfo
  {year} {2006})%
  \bibAnnoteFile{NoStop}{Bertl:2006up}%
\bibitem{Blondel:2013ia}%
  \BibitemOpen
  \bibfield{author}{%
  \bibinfo {author} {\bibfnamefont{A.}~\bibnamefont{Blondel}} \emph{et~al.},\
  }%
  \bibfield{title}{%
  \enquote{\bibinfo {title} {{Research Proposal for an Experiment to Search for
  the Decay $\mu \to eee$}},}\ }%
   (\bibinfo {month} {1}\ \bibinfo {year} {2013}),\
  \Eprint{http://arxiv.org/abs/1301.6113}{arXiv:1301.6113 [physics.ins-det]}%
  \bibAnnoteFile{NoStop}{Blondel:2013ia}%
\bibitem{Chiang:2019ajm}%
  \BibitemOpen
  \bibfield{author}{%
  \bibinfo {author} {\bibfnamefont{Cheng-Wei}\ \bibnamefont{Chiang}}, \bibinfo
  {author} {\bibfnamefont{Giovanna}\ \bibnamefont{Cottin}}, \bibinfo {author}
  {\bibfnamefont{Arindam}\ \bibnamefont{Das}},\ and\ \bibinfo {author}
  {\bibfnamefont{Sanjoy}\ \bibnamefont{Mandal}},\ }%
  \bibfield{title}{%
  \enquote{\bibinfo {title} {{Displaced heavy neutrinos from $Z'$ decays at the
  LHC}},}\ }%
  \bibfield{journal}{%
  \Doi{10.1007/JHEP12(2019)070}{\bibinfo {journal} {JHEP}}\ }%
  \textbf{\bibinfo {volume} {12}},\ \bibinfo {pages} {070} (\bibinfo {year}
  {2019}),\ \Eprint{http://arxiv.org/abs/1908.09838}{arXiv:1908.09838
  [hep-ph]}%
  \bibAnnoteFile{NoStop}{Chiang:2019ajm}%
\bibitem{Aad:2019fac}%
  \BibitemOpen
  \bibfield{author}{%
  \bibinfo {author} {\bibfnamefont{Georges}\ \bibnamefont{Aad}} \emph{et~al.}
  (\bibinfo {collaboration} {ATLAS}),\ }%
  \bibfield{title}{%
  \enquote{\bibinfo {title} {{Search for high-mass dilepton resonances using
  139 fb$^{-1}$ of $pp$ collision data collected at $\sqrt{s}=$13 TeV with the
  ATLAS detector}},}\ }%
  \bibfield{journal}{%
  \Doi{10.1016/j.physletb.2019.07.016}{\bibinfo {journal} {Phys. Lett. B}}\ }%
  \textbf{\bibinfo {volume} {796}},\ \bibinfo {pages} {68--87} (\bibinfo {year}
  {2019}),\ \Eprint{http://arxiv.org/abs/1903.06248}{arXiv:1903.06248
  [hep-ex]}%
  \bibAnnoteFile{NoStop}{Aad:2019fac}%
\bibitem{Aad:2019vnb}%
  \BibitemOpen
  \bibfield{author}{%
  \bibinfo {author} {\bibfnamefont{Georges}\ \bibnamefont{Aad}} \emph{et~al.}
  (\bibinfo {collaboration} {ATLAS}),\ }%
  \bibfield{title}{%
  \enquote{\bibinfo {title} {{Search for electroweak production of charginos
  and sleptons decaying into final states with two leptons and missing
  transverse momentum in $\sqrt{s}=13$ TeV $pp$ collisions using the ATLAS
  detector}},}\ }%
  \bibfield{journal}{%
  \Doi{10.1140/epjc/s10052-019-7594-6}{\bibinfo {journal} {Eur. Phys. J. C}}\
  }%
  \textbf{\bibinfo {volume} {80}},\ \bibinfo {pages} {123} (\bibinfo {year}
  {2020}),\ \Eprint{http://arxiv.org/abs/1908.08215}{arXiv:1908.08215
  [hep-ex]}%
  \bibAnnoteFile{NoStop}{Aad:2019vnb}%
\bibitem{Sirunyan:2020eab}%
  \BibitemOpen
  \bibfield{author}{%
  \bibinfo {author} {\bibfnamefont{Albert~M}\ \bibnamefont{Sirunyan}}
  \emph{et~al.} (\bibinfo {collaboration} {CMS}),\ }%
  \bibfield{title}{%
  \enquote{\bibinfo {title} {{Search for supersymmetry in final states with two
  oppositely charged same-flavor leptons and missing transverse momentum in
  proton-proton collisions at $\sqrt{s}=$ 13 TeV}},}\ }%
   (\bibinfo {month} {12}\ \bibinfo {year} {2020}),\
  \Eprint{http://arxiv.org/abs/2012.08600}{arXiv:2012.08600 [hep-ex]}%
  \bibAnnoteFile{NoStop}{Sirunyan:2020eab}%
\bibitem{Crivellin:2020klg}%
  \BibitemOpen
  \bibfield{author}{%
  \bibinfo {author} {\bibfnamefont{Andreas}\ \bibnamefont{Crivellin}}, \bibinfo
  {author} {\bibfnamefont{Fiona}\ \bibnamefont{Kirk}}, \bibinfo {author}
  {\bibfnamefont{Claudio~Andrea}\ \bibnamefont{Manzari}},\ and\ \bibinfo
  {author} {\bibfnamefont{Luca}\ \bibnamefont{Panizzi}},\ }%
  \bibfield{title}{%
  \enquote{\bibinfo {title} {{Searching for Lepton Flavour (Universality)
  Violation and Collider Signals from a Singly-Charged Scalar Singlet}},}\ }%
   (\bibinfo {month} {12}\ \bibinfo {year} {2020}),\
  \Eprint{http://arxiv.org/abs/2012.09845}{arXiv:2012.09845 [hep-ph]}%
  \bibAnnoteFile{NoStop}{Crivellin:2020klg}%
\bibitem{Alcaide:2019kdr}%
  \BibitemOpen
  \bibfield{author}{%
  \bibinfo {author} {\bibfnamefont{Julien}\ \bibnamefont{Alcaide}}\ and\
  \bibinfo {author} {\bibfnamefont{Nicol\'as~I.}\ \bibnamefont{Mileo}},\ }%
  \bibfield{title}{%
  \enquote{\bibinfo {title} {{LHC sensitivity to singly-charged scalars
  decaying into electrons and muons}},}\ }%
  \bibfield{journal}{%
  \Doi{10.1103/PhysRevD.102.075030}{\bibinfo {journal} {Phys. Rev. D}}\ }%
  \textbf{\bibinfo {volume} {102}},\ \bibinfo {pages} {075030} (\bibinfo {year}
  {2020}),\ \Eprint{http://arxiv.org/abs/1906.08685}{arXiv:1906.08685
  [hep-ph]}%
  \bibAnnoteFile{NoStop}{Alcaide:2019kdr}%
\bibitem{Cao:2017ffm}%
  \BibitemOpen
  \bibfield{author}{%
  \bibinfo {author} {\bibfnamefont{Qing-Hong}\ \bibnamefont{Cao}}, \bibinfo
  {author} {\bibfnamefont{Gang}\ \bibnamefont{Li}}, \bibinfo {author}
  {\bibfnamefont{Ke-Pan}\ \bibnamefont{Xie}},\ and\ \bibinfo {author}
  {\bibfnamefont{Jue}\ \bibnamefont{Zhang}},\ }%
  \bibfield{title}{%
  \enquote{\bibinfo {title} {{Searching for Weak Singlet Charged Scalar at the
  Large Hadron Collider}},}\ }%
  \bibfield{journal}{%
  \Doi{10.1103/PhysRevD.97.115036}{\bibinfo {journal} {Phys. Rev. D}}\ }%
  \textbf{\bibinfo {volume} {97}},\ \bibinfo {pages} {115036} (\bibinfo {year}
  {2018}),\ \Eprint{http://arxiv.org/abs/1711.02113}{arXiv:1711.02113
  [hep-ph]}%
  \bibAnnoteFile{NoStop}{Cao:2017ffm}%
\bibitem{Babu:2019mfe}%
  \BibitemOpen
  \bibfield{author}{%
  \bibinfo {author} {\bibfnamefont{K.~S.}\ \bibnamefont{Babu}}, \bibinfo
  {author} {\bibfnamefont{P.~S.~Bhupal}\ \bibnamefont{Dev}}, \bibinfo {author}
  {\bibfnamefont{Sudip}\ \bibnamefont{Jana}},\ and\ \bibinfo {author}
  {\bibfnamefont{Anil}\ \bibnamefont{Thapa}},\ }%
  \bibfield{title}{%
  \enquote{\bibinfo {title} {{Non-Standard Interactions in Radiative Neutrino
  Mass Models}},}\ }%
  \bibfield{journal}{%
  \Doi{10.1007/JHEP03(2020)006}{\bibinfo {journal} {JHEP}}\ }%
  \textbf{\bibinfo {volume} {03}},\ \bibinfo {pages} {006} (\bibinfo {year}
  {2020}),\ \Eprint{http://arxiv.org/abs/1907.09498}{arXiv:1907.09498
  [hep-ph]}%
  \bibAnnoteFile{NoStop}{Babu:2019mfe}%
\bibitem{Alcaide:2017dcx}%
  \BibitemOpen
  \bibfield{author}{%
  \bibinfo {author} {\bibfnamefont{Julien}\ \bibnamefont{Alcaide}}, \bibinfo
  {author} {\bibfnamefont{Mikael}\ \bibnamefont{Chala}},\ and\ \bibinfo
  {author} {\bibfnamefont{Arcadi}\ \bibnamefont{Santamaria}},\ }%
  \bibfield{title}{%
  \enquote{\bibinfo {title} {{LHC signals of radiatively-induced neutrino
  masses and implications for the Zee\textendash{}Babu model}},}\ }%
  \bibfield{journal}{%
  \Doi{10.1016/j.physletb.2018.02.001}{\bibinfo {journal} {Phys. Lett. B}}\ }%
  \textbf{\bibinfo {volume} {779}},\ \bibinfo {pages} {107--116} (\bibinfo
  {year} {2018}),\ \Eprint{http://arxiv.org/abs/1710.05885}{arXiv:1710.05885
  [hep-ph]}%
  \bibAnnoteFile{NoStop}{Alcaide:2017dcx}%
\bibitem{Aad:2014hja}%
  \BibitemOpen
  \bibfield{author}{%
  \bibinfo {author} {\bibfnamefont{Georges}\ \bibnamefont{Aad}} \emph{et~al.}
  (\bibinfo {collaboration} {ATLAS}),\ }%
  \bibfield{title}{%
  \enquote{\bibinfo {title} {{Search for new phenomena in events with three or
  more charged leptons in $pp$ collisions at $\sqrt{s}=8$ TeV with the ATLAS
  detector}},}\ }%
  \bibfield{journal}{%
  \Doi{10.1007/JHEP08(2015)138}{\bibinfo {journal} {JHEP}}\ }%
  \textbf{\bibinfo {volume} {08}},\ \bibinfo {pages} {138} (\bibinfo {year}
  {2015}),\ \Eprint{http://arxiv.org/abs/1411.2921}{arXiv:1411.2921 [hep-ex]}%
  \bibAnnoteFile{NoStop}{Aad:2014hja}%
\bibitem{Rodejohann:2017vup}%
  \BibitemOpen
  \bibfield{author}{%
  \bibinfo {author} {\bibfnamefont{Werner}\ \bibnamefont{Rodejohann}}, \bibinfo
  {author} {\bibfnamefont{Xun-Jie}\ \bibnamefont{Xu}},\ and\ \bibinfo {author}
  {\bibfnamefont{Carlos~E.}\ \bibnamefont{Yaguna}},\ }%
  \bibfield{title}{%
  \enquote{\bibinfo {title} {{Distinguishing between Dirac and Majorana
  neutrinos in the presence of general interactions}},}\ }%
  \bibfield{journal}{%
  \Doi{10.1007/JHEP05(2017)024}{\bibinfo {journal} {JHEP}}\ }%
  \textbf{\bibinfo {volume} {05}},\ \bibinfo {pages} {024} (\bibinfo {year}
  {2017}),\ \Eprint{http://arxiv.org/abs/1702.05721}{arXiv:1702.05721
  [hep-ph]}%
  \bibAnnoteFile{NoStop}{Rodejohann:2017vup}%
\bibitem{Bellini:2011rx}%
  \BibitemOpen
  \bibfield{author}{%
  \bibinfo {author} {\bibfnamefont{G.}~\bibnamefont{Bellini}} \emph{et~al.},\
  }%
  \bibfield{title}{%
  \enquote{\bibinfo {title} {{Precision measurement of the 7Be solar neutrino
  interaction rate in Borexino}},}\ }%
  \bibfield{journal}{%
  \Doi{10.1103/PhysRevLett.107.141302}{\bibinfo {journal} {Phys. Rev. Lett.}}\
  }%
  \textbf{\bibinfo {volume} {107}},\ \bibinfo {pages} {141302} (\bibinfo {year}
  {2011}),\ \Eprint{http://arxiv.org/abs/1104.1816}{arXiv:1104.1816 [hep-ex]}%
  \bibAnnoteFile{NoStop}{Bellini:2011rx}%
\bibitem{Deniz:2009mu}%
  \BibitemOpen
  \bibfield{author}{%
  \bibinfo {author} {\bibfnamefont{M.}~\bibnamefont{Deniz}} \emph{et~al.}
  (\bibinfo {collaboration} {TEXONO}),\ }%
  \bibfield{title}{%
  \enquote{\bibinfo {title} {{Measurement of Nu(e)-bar -Electron Scattering
  Cross-Section with a CsI(Tl) Scintillating Crystal Array at the Kuo-Sheng
  Nuclear Power Reactor}},}\ }%
  \bibfield{journal}{%
  \Doi{10.1103/PhysRevD.81.072001}{\bibinfo {journal} {Phys. Rev. D}}\ }%
  \textbf{\bibinfo {volume} {81}},\ \bibinfo {pages} {072001} (\bibinfo {year}
  {2010}),\ \Eprint{http://arxiv.org/abs/0911.1597}{arXiv:0911.1597 [hep-ex]}%
  \bibAnnoteFile{NoStop}{Deniz:2009mu}%
\bibitem{Vilain:1994qy}%
  \BibitemOpen
  \bibfield{author}{%
  \bibinfo {author} {\bibfnamefont{P.}~\bibnamefont{Vilain}} \emph{et~al.}
  (\bibinfo {collaboration} {CHARM-II}),\ }%
  \bibfield{title}{%
  \enquote{\bibinfo {title} {{Precision measurement of electroweak parameters
  from the scattering of muon-neutrinos on electrons}},}\ }%
  \bibfield{journal}{%
  \Doi{10.1016/0370-2693(94)91421-4}{\bibinfo {journal} {Phys. Lett. B}}\ }%
  \textbf{\bibinfo {volume} {335}},\ \bibinfo {pages} {246--252} (\bibinfo
  {year} {1994})%
  \bibAnnoteFile{NoStop}{Vilain:1994qy}%
\bibitem{Porod:2003um}%
  \BibitemOpen
  \bibfield{author}{%
  \bibinfo {author} {\bibfnamefont{Werner}\ \bibnamefont{Porod}},\ }%
  \bibfield{title}{%
  \enquote{\bibinfo {title} {{SPheno, a program for calculating supersymmetric
  spectra, SUSY particle decays and SUSY particle production at e+ e-
  colliders}},}\ }%
  \bibfield{journal}{%
  \Doi{10.1016/S0010-4655(03)00222-4}{\bibinfo {journal} {Comput. Phys.
  Commun.}}\ }%
  \textbf{\bibinfo {volume} {153}},\ \bibinfo {pages} {275--315} (\bibinfo
  {year} {2003}),\
  \Eprint{http://arxiv.org/abs/hep-ph/0301101}{arXiv:hep-ph/0301101 [hep-ph]}%
  \bibAnnoteFile{NoStop}{Porod:2003um}%
\bibitem{Porod:2011nf}%
  \BibitemOpen
  \bibfield{author}{%
  \bibinfo {author} {\bibfnamefont{W.}~\bibnamefont{Porod}}\ and\ \bibinfo
  {author} {\bibfnamefont{F.}~\bibnamefont{Staub}},\ }%
  \bibfield{title}{%
  \enquote{\bibinfo {title} {{SPheno 3.1: Extensions including flavour,
  CP-phases and models beyond the MSSM}},}\ }%
  \bibfield{journal}{%
  \Doi{10.1016/j.cpc.2012.05.021}{\bibinfo {journal} {Comput. Phys. Commun.}}\
  }%
  \textbf{\bibinfo {volume} {183}},\ \bibinfo {pages} {2458--2469} (\bibinfo
  {year} {2012}),\ \Eprint{http://arxiv.org/abs/1104.1573}{arXiv:1104.1573
  [hep-ph]}%
  \bibAnnoteFile{NoStop}{Porod:2011nf}%
\bibitem{Staub:2013tta}%
  \BibitemOpen
  \bibfield{author}{%
  \bibinfo {author} {\bibfnamefont{Florian}\ \bibnamefont{Staub}},\ }%
  \bibfield{title}{%
  \enquote{\bibinfo {title} {{SARAH 4 : A tool for (not only SUSY) model
  builders}},}\ }%
  \bibfield{journal}{%
  \Doi{10.1016/j.cpc.2014.02.018}{\bibinfo {journal} {Comput. Phys. Commun.}}\
  }%
  \textbf{\bibinfo {volume} {185}},\ \bibinfo {pages} {1773--1790} (\bibinfo
  {year} {2014}),\ \Eprint{http://arxiv.org/abs/1309.7223}{arXiv:1309.7223
  [hep-ph]}%
  \bibAnnoteFile{NoStop}{Staub:2013tta}%
\bibitem{Staub:2015kfa}%
  \BibitemOpen
  \bibfield{author}{%
  \bibinfo {author} {\bibfnamefont{Florian}\ \bibnamefont{Staub}},\ }%
  \bibfield{title}{%
  \enquote{\bibinfo {title} {{Exploring new models in all detail with
  SARAH}},}\ }%
  \bibfield{journal}{%
  \Doi{10.1155/2015/840780}{\bibinfo {journal} {Adv. High Energy Phys.}}\ }%
  \textbf{\bibinfo {volume} {2015}},\ \bibinfo {pages} {840780} (\bibinfo
  {year} {2015}),\ \Eprint{http://arxiv.org/abs/1503.04200}{arXiv:1503.04200
  [hep-ph]}%
  \bibAnnoteFile{NoStop}{Staub:2015kfa}%
\end{thebibliography}%

\end{document}